\documentclass[english,showpacs,floatfix,preprint,10pt]{revtex4}
\usepackage[dvips]{graphicx}
\usepackage{longtable}
\usepackage{amsmath,amssymb}
\usepackage{dcolumn}
\usepackage{latexsym}
\usepackage{babel}
\usepackage[latin1]{inputenc}
\usepackage{color}
\usepackage[colorlinks]{hyperref}

\begin{document}
\baselineskip=0.8 cm
\title{{\bf The imprint of the interaction between dark sectors in galaxy clusters}}

\author{Jian-Hua He, Bin Wang}   \affiliation{INPAC and Department of Physics, Shanghai Jiao Tong University, 200240 Shanghai, China}
\author{Elcio Abdalla}
\affiliation{Instituto de Fisica, Universidade de Sao Paulo, CP 66318, 05315-970, Sao Paulo, Brazil}
\author{Diego Pavon} \affiliation{Department of Physics, Autonomous University of Barcelona, 08193 Bellaterra, Barcelona, Spain}

\vspace*{0.2cm}
\begin{abstract}
\baselineskip=0.6 cm

Based on perturbation theory, we study the dynamics of how dark matter and dark energy in the collapsing system approach dynamical equilibrium
when they are in interaction. We find that the interaction between dark sectors cannot ensure the dark energy to fully cluster along with dark
matter. When dark energy does not trace dark matter, we present a new treatment on studying the structure formation in the spherical collapsing
system. Furthermore we examine the cluster number counts dependence on the interaction between dark sectors and analyze how dark energy
inhomogeneities affect cluster abundances. It is shown that cluster number counts can provide specific signature of dark sectors interaction and
dark energy inhomogeneities.

\end{abstract}

 \maketitle
\newpage
\section{introduction}

 There has been convincing evidence indicating that our universe is experiencing an accelerated expansion\cite{1}. In the framework of General  Relativity, the impart of the acceleration of the universe is given by a kind of mysterious energy called dark
energy(DE).  One leading candidate of such DE is the cosmological constant. However, it is difficult to understand such a cosmological constant in terms of fundamental physics. Its value from observation is far below that calculated in quantum field theory, what is referred to as the cosmological constant problem. Moreover, using the cosmological constant to explain the DE, one is unavoidably led to the coincidence problem, namely, why the vacuum and matter energy densities are precisely of the same order today.

Considering that DE contributes a significant fraction of the content of the universe, it is natural in the framework of field theory to consider its interactions with the remaining fields, as e.g., the other large fraction of the universe, the dark matter (DM). The possibility that DE and DM interact with each other has been widely discussed recently \cite{10}-\cite{AbdallaPLB09}. It has been shown that an appropriate interaction between DE and DM can provide a mechanism to alleviate the coincidence problem \cite{10,11,13,Olivares,14}. However, it has been suspected
that the appearance of the coupling between dark sectors may lead to the curvature perturbation instability \cite{maartens}. This problem was later clarified in \cite{hePLB09} where it has been shown that the stability of curvature perturbation holds while appropriately choosing the forms of the interaction between dark sectors and the equations of state (EOS) of DE. Another possibility to cure the instability was suggested in \cite{stability}. Observational signatures on the dark sector mutual interaction have been found in the probes of the cosmic expansion history by
using the WMAP, SNIa, BAO and SDSS data etc \cite{71}-\cite{hePRD09}. Interestingly it was disclosed that the late ISW effect has the unique ability to give insight into the coupling between dark sectors \cite{hePRD09}. The small positive coupling indicates that there is energy transfer from DE to DM, what can help to alleviate the coincidence problem \cite{hePRD09,heJCAP08}.

To further unveil the nature of the interaction between DE and DM, we require complementary probes in addition to investigating the expansion
history of the universe. It has been observed that the growth of cosmic structure can be influenced by a nonminimal interaction between dark sectors, what leaves a clear change in the growth index due to the coupling \cite{31, Caldera}. Furthermore, it was suggested that the dynamical equilibrium of collapsed structures such as clusters would acquire a modification due to the coupling between DE and DM \cite{AbdallaPLB09,pt}. Comparing the naive virial masses of a large sample of clusters with their masses estimated by X-ray and by weak lensing data, small positive coupling has been tightly constrained \cite{AbdallaPLB09}, what agrees with the results given in \cite{hePRD09} from CMB.

The redshift dependence of cluster number counts is another promising tool to discriminate different DE models. Discussions can be found in \cite{multamaki2003}-\cite{Abramo20072009}. Similar investigation to test the particular scenario of coupled quintessence model by using the
cluster number counts was investigated in \cite{David,Nunes,mota2006}, where the interaction between quintessence and DM was specifically described by
some product of the densities of DE and DM.  The sign they used for the coupling between dark sectors just describes the energy decay from DM to
DE and they concentrated on the inhomogeneities in the DM fluid propagating in the scalar field and affecting the scalar field evolution. The
cluster number counts dependence on the amount of DM coupled to DE was analyzed therein.

In this paper we will investigate the possibility of employing measurements from cluster number counts to grasp the signature of interactions
between DE and DM in some commonly discussed phenomenological models. These models were proved to have stable curvature perturbations
\cite{hePLB09} and have been tested by confronting CMB \cite{hePRD09} and virial mass of cluster observations \cite{AbdallaPLB09}. In our study
we will use the spherical collapse model and consider either DE being distributed homogeneously or inhomogeneously.  We will discuss the dynamics
of DM and the inhomogeneous DE to virialization. When DE is homogeneous, DM flows to equilibrium in agreement with that discussed in
\cite{Layzer} if there is no interaction between dark sectors. If the DE is inhomogeneous, we will show that its dynamics is different from that
of DM in the virialization in the linear perturbation level. This indicates that DE does not cluster fully along with the DM and thus energy is
not strictly conserved inside the collapsing system in the linear perturbation theory. Our result is a progress in the linear perturbation
formalism, which has not been reported before.

In the following discussion of this work, we will use the Press and Schechter formalism \cite{ps} to predict the number density of collapsed
objects. We will develop a new  treatment on studying how the structure is formed when DE does not trace DM.  The Press and Schechter
formalism is crude if compared with N-body simulations, while here we do not seek the precise confrontations with the observational data, but the
evolution of the mass function of collapsed objects predicted by the formalism will be helpful for us to understand the influence of the
interaction between dark sectors and the inhomogeneous DE distribution on cluster number counts.

This paper is organized as follows: In Section~\ref{perturbation}, we review the formalism of the perturbation theory in the presence of
interaction between dark sectors. In Section~\ref{Layer}, we study the dynamics to describe the flow of the collapsing object to virialization.
In Section~\ref{spherical}, we introduce the spherical collapse model in the presence of coupling between DE and DM and the Press-Schechter
formalism for the galaxy number counts. In Section~\ref{numerical}, we show the numerical results on how the interaction and inhomogeneous DE
influence the cluster number counts. In the last section we present our summary and discussions.

\section{PERTURBATION THEORY WHEN DE INTERACTS WITH DM \label{perturbation}}
In this section, we go over the main results on the linear perturbation theory when  DE interacts with DM. The detailed descriptions can be found
in \cite{hePLB09,31}.

In the spatially flat Friedmann-Robertson-Walker(FRW) background, if there is interaction between DE and DM, neither of them can evolve
independently. The (non)conservation equation are
\begin{eqnarray}
&&\rho'_m+3\mathcal{H}\rho_m=a^2Q_m^0,\nonumber \\
&&\rho'_d+3\mathcal{H}(1+w)\rho_d=a^2Q_d^0,\label{friedmann}
\end{eqnarray}
where the subscript ``m" denotes the cold DM, ``d" denotes the DE and $Q$ describes the interaction between them. Since we know neither the physics of DM nor that of DE at the present moment, we cannot write out the precise form of the
interaction between them from first principles (see \cite{maw} for recent attempts). One has to specify the interaction either from the
outset\cite{11}, or determine it from phenomenological requirements \cite{heJCAP08,Olivares}. For the sake of generality, we consider the
phenomenological description of the interaction between DE and DM in the comoving frame \cite{hePLB09,31}
\begin{eqnarray}
Q_m^{\nu}&=&\left[\frac{3\mathcal{H}}{a^2}(\xi_1\rho_m+\xi_2\rho_d),0,0,0\right]^{T}\nonumber\\
Q_d^{\nu}&=&\left[-\frac{3\mathcal{H}}{a^2}(\xi_1\rho_m+\xi_2\rho_d),0,0,0\right]^{T},\label{couplingvector}
\end{eqnarray}
where $\mathcal{H}$ is the global expansion rate in the conformal time, $\xi_1, \xi_2$ are small dimensionless constants and $T$ is the
transpose of the matrix.

The perturbation equations for DM and DE in the subhorizon scale have been derived in \cite{31}, having the form
\begin{eqnarray}
\Delta_m'&=& -kV_m-\Delta_m\frac{a^2Q_m^0}{\rho_m}+\frac{a^2\delta Q^{0I}_m}{\rho_m},\nonumber \\
V_m'&=&-\mathcal{H}V_m+k\Psi-\frac{a^2Q_m^0}{\rho_m}V_m+\frac{a^2\delta Q_{pm}^I}{\rho_m};\\
\Delta_d'&=& 3\mathcal{H}(w-C_e^2)\Delta_d-k(1+w)V_d-\Delta_d\frac{a^2Q^{0}_d}{\rho_d}+\frac{a^2\delta Q^{0I}_d}{\rho_d},\nonumber \\
V_d'&+&\mathcal{H}(1-3w)V_d=\frac{kC_e^2}{1+w} \Delta_d+\frac{C_a^2}{1+w}\frac{\rho_d'}{\rho_d}V_d+k
\Psi-\frac{w'}{1+w}V_d-\frac{a^2Q_d^0}{\rho_d}V_d+\frac{a^2\delta Q_{pd}^I}{(1+w)\rho_d};
\end{eqnarray}
where the prime denotes the derivative with respect to the conformal time and the gauge-invariant quantities $\Delta_m,\Delta_d$ represent the
density contrast $\Delta_m\approx \delta \rho_m/\rho_m=\delta_m$, $\Delta_d\approx \delta \rho_d/\rho_d=\delta_d$. The perturbed gauge invariant couplings are
expressed as \cite{hePLB09,31},
$$\frac{a^2\delta Q^{0I}_m}{\rho_m}\approx 3\mathcal{H}(\xi_1\Delta_m+\xi_2\Delta_d/r),$$
$$\frac{a^2\delta Q^{0I}_d}{\rho_d}\approx -3\mathcal{H}(\xi_1\Delta_mr+\xi_2\Delta_d),$$
where $r=\rho_m/\rho_d$.

It is useful to rewrite these equations in the real space,
\begin{eqnarray}
&&\Delta_m'+\nabla_{\bar{x}}\cdot V_m= 3\mathcal{H}\xi_2(\Delta_d-\Delta_m)/r \quad ,\nonumber \\
&&V_m'+\mathcal{H}V_m=-\nabla_{\bar{x}}\Psi-3\mathcal{H}(\xi_1+\xi_2/r)V_m\quad ; \label{DM1}\\
&&\Delta_d'+(1+w)\nabla_{\bar{x}}\cdot V_d= 3\mathcal{H}(w-C_e^2)\Delta_d+3\mathcal{H}\xi_1r(\Delta_d-\Delta_m)\quad ,\nonumber \\
&&V_d'+\mathcal{H}V_d=-\nabla_{\bar{x}}\Psi-\frac{C_e^2}{1+w}
\nabla_{\bar{x}}\Delta_d-\frac{w'}{1+w}V_d+3\mathcal{H}
\left\{(w-C_a^2)+\frac{1+w-C_a^2}{1+w}(\xi_1r+\xi_2)\right\}Vd\quad ;\label{DE1}
\end{eqnarray}
where ${\bar{x}}$ refers to the conformal coordinates.

Defining $\sigma_m=\delta \rho_m$, $\sigma_d=\delta \rho_d$, and assuming that the EOS of DE is constant $w'=0$, we can change
Eqs.(~\ref{DM1},~\ref{DE1}) into,
\begin{eqnarray}
&&\dot{\sigma}_m+3H\sigma_m+\nabla_{x}(\rho_mV_m)=3H(\xi_1\sigma_m+\xi_2\sigma_d),\nonumber \\
&&\frac{\partial}{\partial t}(a V_m)=-\nabla_x(a \Psi)-3H(\xi_1+\xi_2/r)(aV_m);\label{dmn}\\
&&\dot{\sigma}_d+3H(1+C_e^2)\sigma_d+(1+w)\nabla_{x}(\rho_dV_d)=-3H(\xi_1\sigma_m+\xi_2 \sigma_d),\nonumber \\
&&\frac{\partial}{\partial t}(aV_d)=-\nabla_x(a \Psi)-\frac{C_e^2}{1+w}\nabla_x\cdot (a
\Delta_d)+3H\left[(w-c_a^2)+\frac{1+w-c_a^2}{1+w}(\xi_1r+\xi_2)\right](aV_d);\label{den}
\end{eqnarray}
where $\nabla_{x}=\frac{1}{a}\nabla_{\bar{x}}$. The dot denotes the derivative with respect to the cosmic time. $\Psi$ indicates the peculiar
potential, which can be decomposed into $\Psi=\psi_m+\psi_d$, satisfying the Poisson  equation \cite{Motar04},
\begin{equation}
\nabla^2_{\lambda}\psi_\lambda=4\pi G (1+3w_\lambda)\sigma_\lambda,
\end{equation}
where $\sigma_{\lambda}$ represents the inhomogeneous fluctuation field and the subscript "$\lambda$" denotes DM or DE, respectively. We have
included the correction from General Relativity. In a homogeneous and isotropic background $<\psi_\lambda>=0$, since
$<\sigma_\lambda>=0$. For DE and DM, their peculiar potentials read \cite{Motar04}
\begin{equation}
\psi_{m}=-4\pi G\int dV'\frac{\sigma_{m}}{\mid x-x' \mid},
\end{equation}
\begin{equation}
\psi_{d}=-4\pi G\int dV' \frac{(1+3w)\sigma_{d}}{\mid x-x' \mid}.
\end{equation}

\section{Derivation of the Layzer-Irvine equation \label{Layer}}

In this section we derive the Layer-Irvine equation \cite{Layzer} when there is an interaction between DE and DM. Layzer-Irvine equation describes
how a collapsing system reaches a state of dynamical equilibrium in an expanding universe. In the presence of the interaction between DE and DM, the
Layer-Irvine equation can tell us how the coupling between DE and DM affects the process of attaining equilibrium as well as the respective configuration. Our derivation will be based on the perturbation theory given in Sec.~\ref{perturbation}.

For DM, the rate of change of the peculiar velocity is given by Eqs.(\ref{dmn})
\begin{equation}
\frac{\partial}{\partial t}(a V_m)=-\nabla_x(a \psi_m+ a\psi_d)-3H(\xi_1+\xi_2/r)(aV_m).\label{dmn1}
\end{equation}
Neglecting the influence of DE and the couplings, the above equation is nothing but the rate of change of the peculiar velocity of the DM
particle in the expanding universe described by the Newton's law which was the starting point in \cite{Layzer}. To derive the energy equation for
local irregularities, we follow the same procedure as \cite{Layzer} by forming a product of Eqs.(\ref{dmn1}) with $aV_m\rho_m\hat{\varepsilon}$
and integrating over the volume. Here $\hat{\varepsilon}=a^3d\bar{x}\wedge d\bar{y}\wedge d\bar{z}$ is the volume element which satisfies
$\frac{\partial}{\partial t}\hat{\varepsilon}=3H\hat{\varepsilon}$. Considering Eqs.(\ref{friedmann}), the LHS of Eq.(\ref{dmn1}) can be multiplied by $a V_m$ and integrated to yield
\begin{equation}
\frac{\partial}{\partial t}\left(a^2T_m\right)-a^23H(\xi_1+\xi_2/r)T_m. \label{Tm}
\end{equation}
where $ T_m=\frac{1}{2}\int V_m^2 \rho_m \hat{\varepsilon}$ is the kinetic energy of DM associated with peculiar motions of DM particles.

The RHS of Eqs.(~\ref{dmn1}) can be treated by the same token. Using partial integration, the potential part can be changed into
\begin{equation}
-\int aV_m \nabla_{x}(a \psi_m+a\psi_d) \rho_m \hat{\varepsilon}=a^2 \int \nabla_{x}(\rho_m V_m)\psi_m \hat{\varepsilon}+a^2 \int
\nabla_{x}(\rho_m V_m)\psi_d \hat{\varepsilon}.\nonumber
\end{equation}
Taking account of the first equation in (\ref{dmn}), it can turn into
\begin{eqnarray}
-\int aV_m \nabla_{x}(a \psi_m&+&a\psi_d) \rho_m \hat{\varepsilon}=-a^2(\dot{U}_{mm}+HU_{mm})-a^2\int \psi_d \frac{\partial}{\partial t} (\sigma_m \hat{\varepsilon})\nonumber\\
&+&3a^2H\left\{\xi_1U_{md}+\xi_2U_{dm}+2\xi_1U_{mm}+2\xi_2U_{dd}\right\}\label{firstdm}
\end{eqnarray}
where $U_{mm}=\frac{1}{2}\int \sigma_m \psi_m \hat{\varepsilon} $, $U_{dm}=\int \sigma_d \psi_m \hat{\varepsilon} $ , $U_{md}=\int \sigma_m
\psi_d \hat{\varepsilon} $, and $U_{dd}=\frac{1}{2}\int \sigma_d \psi_d \hat{\varepsilon} $.

The second term in the RHS of Eq.(\ref{dmn1}) can be changed into,
\begin{equation}
-\int(aV_m)^23H(\xi_1+\xi_2/r)\rho_m\hat{\varepsilon}=-a^26H(\xi_1+\xi_2/r)T_m \label{seconddm}
\end{equation}

Combining Eqs.(\ref{Tm},\ref{firstdm},\ref{seconddm}), we obtain
\begin{eqnarray}
\dot{T}_m+\dot{U}_{mm}&+&H(2T_m+U_{mm})=-\int \psi_d\frac{\partial}{\partial t}(\sigma_m \hat{\varepsilon})-3H(\xi_1+\xi_2/r)T_m \nonumber \\
&&+3H\left\{\xi_1U_{md}+\xi_2U_{dm}+2\xi_1U_{mm}+2\xi_2U_{dd}\right\}.\label{DMv}
\end{eqnarray}
This equation describes how DM reaches dynamical equilibrium in the collapsing system in the expanding universe. If the DE is distributed
homogeneously, $\sigma_d=0$, this equation reduces to
\begin{eqnarray}
\dot{T}_m+\dot{U}_{mm}&+&H(2T_m+U_{mm})=-3H(\xi_1+\xi_2/r)T_m +6H\xi_1U_{mm}.
\end{eqnarray}
For a system in equilibrium, $\dot{T}_m=\dot{U}_{mm}=0$, we get the virial condition. If we take $\bar{\xi_1}=3\xi_1, \bar{\xi_2}=3\xi_2$, we
have the virial condition $(2+\bar{\xi_1}+\bar{\xi_2}/r)T_m+(1-2\bar{\xi_1})U_{mm}=0$ \cite{AbdallaPLB09}. Neglecting the interaction
$\bar{\xi_1}=\bar{\xi_2}=0$, we recover the usual virial condition obtained in \cite{Layzer}. One can see that the presence of the coupling
between DE and DM changes both the time required by the system to reach equilibrium and the equilibrium configuration itself \cite{AbdallaPLB09}.

Now we consider the case that DE is no longer homogeneous and has perturbation. The rate of change of the peculiar velocity of DE is described by the second equation in (\ref{den}). Multiplying both sides of this equation by $aV_d\rho_d\hat{\varepsilon}$ and integrating over the volume, on the LHS we have
\begin{equation}
\frac{\partial}{\partial t}\left(a^2T_d\right)+3a^2H(w+\xi_1r+\xi_2)T_d\label{de1} \quad .
\end{equation}
On the RHS the first term reads,
\begin{eqnarray}
-\int aV_d \nabla_{x}(a \psi_m&+&a\psi_d) \rho_d \hat{\varepsilon}=-\frac{a^2}{1+w}(\dot{U}_{dd}+HU_{dd})-\frac{a^2}{1+w}\int \psi_m\frac{\partial}{\partial t}(\sigma_d \hat{\varepsilon})\nonumber \\
&&-\frac{a^2}{1+w}3H\left\{2(C_e^2+\xi_2)U_{dd}+2\xi_1U_{mm}+\xi_1U_{md}+(C_e^2+\xi_2)U_{dm}\right \}\quad .\label{de2}
\end{eqnarray}
For the remaining terms, we have
\begin{eqnarray}
&&-\frac{c_e^2}{1+w}\int aV_d\nabla_x(a \Delta_d)\rho_d \hat{\varepsilon} + 3H\left [(w-c_a^2)+\frac{1+w-C_a^2}{1+w}(\xi_1r+\xi_2)\right]\int(aV_d)^2 \rho_d \hat{\varepsilon}\nonumber \\
&&=-\frac{c_e^2}{1+w}a^2\int V_d\nabla_x(\sigma_d)
\hat{\varepsilon}+6a^2H\left[(w-C_a^2)+
\frac{1+w-c_a^2}{1+w}(\xi_1r+\xi_2)\right]T_d\quad .\label{de3}
\end{eqnarray}
Combining Eqs.(\ref{de1},\ref{de2},\ref{de3}), we arrive at
\begin{eqnarray}
(1+w)\dot{T}_d&+&\dot{U}_{dd}+H[2(1+w)T_d+U_{dd}]=-3H\left\{2(C_e^2+\xi_2)U_{dd}+2\xi_1U_{mm}+\xi_1U_{md}+(C_e^2+\xi_2)U_{dm} \right\}\nonumber \\
&&-\int\psi_m\frac{\partial}{\partial t}(\sigma_d  \hat{\varepsilon})-c_e^2\int V_d\nabla_x(\sigma_d)
\hat{\varepsilon}+3H\left[(1+w)(w-2C_a^2)+(1+w-2C_a^2)(\xi_1r+\xi_2)\right]T_d, \label{DEv}
\end{eqnarray}
describing how the DE reaches dynamical equilibrium in the collapsing system in an expanding universe if it participates in the structure
formation.

Looking at equations (\ref{DMv}) and (\ref{DEv}), we see that in non-interacting case ($\xi_1=\xi_2=0$), when $C_e^2=0$, DE would cluster just
like cold DM. Examples of DE models with this property were investigated in \cite{Paolo}.  In interacting case, we see that for a collapsing
system to reach dynamical equilibrium, the time and dynamics required by DE and DM to reach equilibrium are different. This shows that in the
collapsing system, DE does not fully cluster along with DM. Thus, the energy conservation breaks down inside the collapsing system. It would be
fair to say that this result was obtained in the linear level. It would be more interesting to examine the result in the non-linear perturbation
and obtain clearer dynamics on how DE participates in the structure formation.

\section{spherical collapse model and the Press-Schechter formalism \label{spherical}}
The simplest analytical tool to study the structure formation is the spherical collapse model. The continuity equation for the background DM and
DE energy densities are given by
\begin{eqnarray}
&&\dot{\rho}_m+3H\rho_m=3H(\xi_1\rho_m+\xi_2\rho_d),\nonumber\\
&&\dot{\rho}_d+3H(1+w)\rho_d=-3H(\xi_1\rho_m+\xi_2\rho_d)\quad .\label{BADE}
\end{eqnarray}
Considering now the spherically symmetric region of radius $R$ and with the energy density
$\rho_{\lambda}^{cluster}=\rho_{\lambda}+\sigma_{\lambda}$, where ``$\lambda$'' denotes DM and DE respectively, it will eventually collapse from
its gravitational pull  provided that $\sigma_{\lambda}>0$. The equation of motion for the collapsing model is governed by Raychaudhuri's
equation.
\begin{equation}
\dot\theta=-\frac{1}{3}\theta^2-4\pi G \sum_{\lambda}( \rho_{\lambda} + 3 p_{\lambda})
\end{equation}
where $\theta=3\frac{\dot{R}}{R}$ and Raychaudhuri's equation can be rewritten as,
\begin{equation}
\frac{\ddot{R}}{R}=-\frac{4\pi G}{3} \sum_{\lambda}( \rho_{\lambda} + 3 p_{\lambda}) \label{newtonian}
\end{equation}
where $R$ is the local expansion scale factor which is determined by the total energy density in the spherical model. If the matter distribution
is homogenous in the interior of the spherical region, the local expansion scale factor  $R$ will have the same value inside the spherical region
and from the Birkhoff theorem, the behavior of $R$ is only determined by the interior matter and is not affected by the matter distributed
outside the spherical region.

In the spherically symmetric region, supposing the DE distribution is homogeneous ($\sigma_d=0$) and its evolution is still described in
(\ref{BADE}), the evolution of the DM energy density in the spherical region behaves as
\begin{equation}
\dot{\rho}_{m}^{cluster}+3h\rho_{m}^{cluster}=3H(\xi_1\rho_{m}^{cluster}+\xi_2\rho_d)
\quad ,
\end{equation}
where $h=\dot{R}/R$.

The Raychaudhuri's equation applied to the
spherical region has the form
\begin{equation}
\ddot{R}=-4\pi G[\frac{1}{3}\rho_{m}^{cluster} +(\frac{1}{3}+w)\rho_d]R\quad .\label{newton}
\end{equation}
where $\rho_d$ is back ground DE energy density. Converting the time derivative into the derivative with respect to the scale factor $a$, we can write
$\ddot{R}=(\dot{a})^2\frac{d^2R}{da^2}+\ddot{a}\frac{dR}{da}$ and change the Raychaudhuri's equation into
\begin{equation}
2a^2(1+\frac{1}{r})R^{''}-R^{'}\left[1+(3w+1)/r\right]a=-R\left[(3w+1)/r+\zeta\right] \quad , \label{dynam}
\end{equation}
where $\zeta=\rho_{m}^{cluster}/\rho_m$. Therefore, we have
\begin{equation}
\zeta'=\frac{3}{a}(1-\xi_2/r)\zeta-3\frac{R'}{R}\zeta+\frac{3}{a}\xi_2/r\quad
,\label{zetam}
\end{equation}
where the prime denotes the derivative with respective to the scale factor $a$ of the universe.

In order to solve these equations, we set the initial conditions $R\sim a$ and $R'=1$ at the starting point $z=3200$ which approximately corresponds to the epoch of
matter-radiation equality. At the starting point the cluster is comoving with the background expansion. In company with the spherical model, we investigate the linear perturbation in the subhorizon approximation satisfying \cite{31}
\begin{eqnarray}
\frac{d^2ln\delta_m}{dlna^2}&+&[\frac{1}{2}-
\frac{3}{2}w(1-\Omega_m)]\frac{dln\delta_m}{dlna}+
\left(\frac{dln\delta_m}{dlna}\right)^2=\nonumber \\
&-&(3\xi_1+6\frac{\xi_2}{r})\frac{dln\delta_m}{dlna}-
\frac{3}{r}[\xi_2+3\xi_1\xi_2+3\xi_2^2/r-\xi_2\frac{dlnr}{dlna}+
\xi_2(\frac{dlnH}{dlna}+1)]+\frac{3}{2}\Omega_m\label{dmlinear}
\end{eqnarray}
and combine with (\ref{dynam},\ref{zetam}) by starting from the initial condition $\zeta_i=\delta_{mi}+1$ in Eqs.~(\ref{zetam}) till the point when the spherical
model collapses $R(a_{coll})\approx0$ to obtain the critical over density $\delta_c$ above which objects collapse. This linear extrapolated
density threshold is useful in the following computation on the halo abundances.

If the DE distribution is not homogeneous, the energy content inside the spherical region consists of multi-fluids. When DE does not fully trace
DM, the four velocity of DE and DM are different $u_{(d)}^a\neq u_{(m)}^a$ and
\begin{equation}
u_{(d)}^a=\gamma(u_{(m)}^a + v_d^a)\label{Lorentz_boost}
\end{equation}
where $\gamma=(1-v_d^2)^{-1/2}$ is Lorentz-boost factor and $v_d^a$ is the relative velocity of DE fluid observed by the observer rested on the
DM frame. If DE fully follows DM, $v_d^a=0$. Now we consider the non-comoving perfect fluids \cite{Christos},
\begin{eqnarray}
T_{(m)}^{ab}&=&\rho_{m}u_{(m)}^au_{(m)}^b\nonumber \\
T_{(d)}^{ab}&=&\rho_du_{(d)}^au_{(d)}^b+p_dh_{(d)}^{ab} \label{non_comoving}
\end{eqnarray}
where $h^{ab}=g^{ab}+u^au^b$ is the projection operator \cite{Christos}. Inserting (\ref{Lorentz_boost}) into the second equation of
(\ref{non_comoving}) and  denoting $u_{(m)}^a$ by $u^a$, the energy momentum tensor for DE reads,
\begin{equation}
T_{(d)}^{ab}=\rho_du^au^b+p_dh^{ab}+2u^{a}q^{b}_{(d)}
\end{equation}
where $q^a=(\rho_d + p_d)v_d^a$ is the energy-flux of DE observed by the observer rested in the DM frame. In the above equation, we have already
neglected the second order terms of $v_d^a$ and assumed that the energy-flux velocity is far less than the speed of light
$v_d^a\ll1,\gamma\sim1$. In the spherical model,  we define the top-hat radius as the radius of the boundary of DM. When DE does not trace DM, DE
will not be bounded inside the top-hat radius and will get out of the spherical region. We assume that the leakage of DE is still spherically
symmetric and from the Birkhoff theorem, in the spherical region, the Raychaudhuri's equation takes the same form as Eqs.~(\ref{newtonian}). For
the energy density conservation law, we have,
\begin{equation}
\nabla_aT^{ab}_{(\lambda)}=Q^b_{(\lambda)}
\end{equation}
where $Q^b$ is the coupling vector as illustrated in Eqs.~\ref{couplingvector} and ``$\lambda$" denotes DE and DM respectively. The timelike parts
of the above equation $u_bQ^b_{(\lambda)}$ give
\begin{eqnarray}
\dot{\rho}_{m}^{cluster}+3h\rho_{m}^{cluster}&=&3H(\xi_1\rho_{m}^{cluster}+\xi_2\rho_d^{cluster})\nonumber \\
\dot{\rho}_{d}^{cluster}+3h(1+w)\rho_{d}^{cluster}&=&-\vartheta(1+w)\rho_d^{cluster}-3H(\xi_1\rho_{m}^{cluster}+\xi_2\rho_d^{cluster})\label{decontin}
\end{eqnarray}
where $\vartheta=\nabla_xv_d$. The external term incorporating $\vartheta$ in the DE density evolution indicates the energy loss caused by the
leakage of DE out of the spherical region.

For the spacelike part, only DE has non-zero spatial component and  $h^{a\phantom{b}}_bQ^b_d$ gives
\begin{equation}
\dot{q}^a_{(d)}+4hq^a_{(d)}=0
\end{equation}
where $q^a_{(d)}$ is the dark energy-flux. Assuming that the energy and pressure are distributed homogenously, we obtain,
\begin{equation}
\dot\vartheta+h(1-3w)\vartheta=3H(\xi_1\Gamma+\xi_2)\vartheta\label{vartheta}
\end{equation}
where $\Gamma=\rho^{cluster}_m/\rho^{cluster}_d$and we have already used Eqs.~(\ref{decontin}) and  kept linear order terms of $\vartheta$. From
Eqs.~(\ref{vartheta}), if $\vartheta$ vanishes initially, $\vartheta$ will keep to be zero at all the time during the evolution, DE will fully
trace DM. However, in most cases, even in linear region there is a small difference between $v_d$ and $v_m$ that the initial condition for
$\vartheta$ is non-zero. We take the initial conditions for $\vartheta$ as $\vartheta\sim k(v_d-v_m)\sim-\delta_{mi}/200<0$, which is obtained
from the prediction of linear equation with $k=1{\rm Mpc}^{-1}$ at $z_i=3200$. $\vartheta$ is much smaller than $1$, $|\vartheta|<<1$, and the
negative value of $\vartheta$ means that at the initial moment DM expanded faster than that of DE.

Defining $\zeta_m = \rho_{m}^{cluster}/\rho_m$, $\zeta_d=\rho_{d}^{cluster}/\rho_d$ and converting the time derivative from $\frac{d}{dt}$ to
$\frac{d}{da}$ we have the evolution of DM and DE in the spherical region described by
\begin{eqnarray}
&&\vartheta'+\frac{R'}{R}(1-3w)\vartheta=\frac{3}{a}(\xi_1\Gamma + \xi_2)\vartheta,\nonumber\\
&&\zeta_m'=\frac{3}{a}\left\{1-\xi_2/r\right\}\zeta_m-3\frac{R'}{R}\zeta_m+\frac{3}{a}\xi_2\zeta_d/r,\nonumber \\
&&\zeta_d'=\frac{3}{a}(1+w+\xi_1r)\zeta_d-3(1+w)\frac{R'}{R}\zeta_d-\frac{3}{a}\xi_1\zeta_mr-\vartheta(1+w)\zeta_d.\label{zetamd}
\end{eqnarray}
When DE fully traces along DM, $\vartheta=0$, only the last two equations above are needed to discuss the structure formation. The Raychaudhuri's
equation now becomes
\begin{equation}
2a^2(1+\frac{1}{r})R^{''}-R^{'}\left[1+(3w+1)/r\right]a=-R\left[(3w+1)\zeta_d/r+\zeta_m\right].\label{dymd}
\end{equation}
Taking the same initial conditions as above for the spherical region of radius $R$ and in addition adopting the adiabatic initial conditions,
$\delta_{di}=(1+w)\delta_{mi}$, and $\zeta_{mi}=\delta_{mi}+1, \zeta_{di}=\delta_{di}+1$ which lead to $\zeta_{di}=(1+w)\zeta_{mi}-w$, we can
study the spherical collapse of  DM and DE . However, it is important to note that due to the coupling, $\zeta_d$ may get negative at sometime
during the collapse. To avoid this unphysical point, we make a cut-off of the coupling when $\zeta_d$ becomes negative to guarantee $\zeta_d>=0$
in our analysis. For linear perturbation, we adopt the equations in the subhorizon approximation through  \cite{31}
\begin{eqnarray}
&&\frac{d^2ln \delta_m}{dlna^2}  = -\left( \frac{dln \delta_m }{d lna }\right)^2 - \left [ \frac{1}{2} - \frac{3}{2}w (1 - \Omega_m)\right]
\frac{dln \delta_m}{d lna}
 -(3\xi_1 +
6\frac{\xi_2}{r})\frac{dln \delta_m}{dlna}+3\frac{\xi_2}{r}\frac{dln \delta_d}{dlna} exp(ln
\frac{\delta_d}{ \delta _m})\nonumber\\
&&+\frac{3[exp (ln \frac{\delta _d}{ \delta _m})-1]}{r}\left\{\xi_2+3\xi_1\xi_2+3\xi_2^2/r+\xi_2(\frac{dln H}{dln a}+1)- \xi_2\frac{dln r}{dln
a}\right\}+\frac{3}{2}\left[\Omega_m+(1 - \Omega_m)exp (ln \frac{\delta _d}{ \delta _m})\right]\quad  ,\label{DMevolution}
\end{eqnarray}
\begin{eqnarray}
&&\frac{d^2ln\delta_d}{dlna^2}= -\left(\frac{dln \delta_d}{dlna}\right)^2 - \left[\frac{1}{2} - \frac{3}{2}w(1-\Omega_m)\right]\frac{dln
\delta_d}{d lna} + (1 + w)\frac{3}{2}\left[\Omega_mexp(ln\frac{\delta_m}{\delta_d})+
(1 - \Omega_m)\right]-\frac{k^2C_e^2}{a^2H^2}\nonumber \\
&+&\left[3\xi_2+6\xi_1r+6w-3C_a^2+3(C_e^2-C_a^2)\frac{\xi_1r+\xi_2}{1+w}+\frac{C_e^2}{1+w}\frac{dln\rho_d}{dln
a}\right]\frac{dln\delta_d}{dlna}\nonumber
\\
&-&3r\xi_1 exp(ln\frac{\delta_m}{\delta_d})\frac{dln\delta_m}{dlna}+3(\frac{dln H}{dln a}+1)(w-C_e^2)
+3\xi_1(\frac{dln H}{dln a}r+r+\frac{dr}{dlna })- 3\xi_1\left[(\frac{dln H}{dln a}+1)r+\frac{d r}{dln a}\right]exp(ln\frac{\delta_m}{\delta_d})\nonumber \\
&+&3\left[w - C_e^2 + \xi_1r(1-
exp(ln\frac{\delta_m}{\delta_d}))\right]\left[(1-3w)-3\frac{C_e^2-C_a^2}{1+w}(1+w+\xi_1r+\xi_2)-3(\xi_1r+\xi_2)-\frac{C_e^2}{1+w}\frac{dln\rho_d}{dln
a}\right]\quad .\label{DEevolution}
\end{eqnarray}
We see that the DE and DM perturbations are entangled. In the subhorizon approximation, it was observed that the influence of the DE perturbation
is small if compared with that DM \cite{31}. Using (~\ref{DMevolution},~\ref{DEevolution}) together with (~\ref{zetamd},~\ref{dymd}), we can obtain in the spherical model with inhomogeneous DE
distribution the linearly extrapolated density threshold, $\delta_c(z)=\delta_m(z=z_{coll})$, above which structure collapse.

The spherical collapse model was used by Press and Schechter\cite{ps} in providing a formalism to predict the number density of collapsed
objects. Although this formalism is crude as found in \cite{Jenlins2001} etc., it predicts the evolution of the mass function of collapsed
objects well enough for the purpose in this paper to see how the interaction between DE and DM influences cluster number counts.

The comoving number density of collapsed dark halos of mass $M$ in the mass interval $dM$ at a given redshift of collapse is given by \cite{ps}
\begin{equation}
\frac{dn(M,z)}{dM}=\sqrt{\frac{2}{\pi}}\frac{\overline{\rho}_m}{3M^2}
\frac{\delta_c}{\sigma}e^{-\delta_c^2/2\sigma^2}
\left[-\frac{R}{\sigma}\frac{d\sigma}{dR}\right]\quad ,\label{probabilities}
\end{equation}
where $\overline{\rho}_m$ is the comoving mean matter density at a given redshift. In most cases it is a constant and equals to the present mean
matter density, but this is not true when DE interacts with DM \cite{mota2006}. The quantity $\sigma=\sigma(R,z)$ here is the
variance smoothed over radius $R$. It has an explicit form \cite{viana},
\begin{equation}
\sigma(R,z)=\sigma_8\left(\frac{R}{8h^{-1}{\rm Mpc}}\right)^{-\gamma(R)}D(z),
\end{equation}
where $\sigma_8$ is the variance over a sphere with radius $R=8h^{-1}{\rm Mpc}$ and $D(z)$ is the growth function defined by
$D(z)=\delta_m(z)/\delta_m(0)$. The index $\gamma$ is a function of the mass scale and the shape parameter $\Gamma$ of the matter power spectrum
\cite{viana}
\begin{equation}
\gamma(R)=(0.3\Gamma +0.2)\left[2.92+\log_{10}\left(\frac{R}{8h^{-1}{\rm Mpc}}\right)\right].
\end{equation}
We will use $\Gamma = 0.3$ throughout our analysis. The radius $R$ at given $M$ can be calculated by the relation \cite{viana},
\begin{equation}
R = 0.951h^{-1}{\rm Mpc}\left (\frac{Mh}{10^{12}\overline{\rho}_m/\rho_c^0M_{s}}\right)^{1/3},
\end{equation}
where $\rho_c^0$ is critical density at present and $M_{s}$ is the solar mass.

The Press-Schechter formalism presents us the comoving number density of halos, which can be compared with astronomical data. In order to do the
comparison, we calculate the all sky number of halos per unit of redshift in the mass bin
\begin{equation}
\frac{dN}{dz}=\int d\Omega\frac{dV}{dzd\Omega}\int n(M)dM,
\end{equation}
where the comoving volume element per unit redshift is $dV/dzd\Omega=r^2(z)/H(z)$ and $r(z)$ is the comoving distance $r(z)=\int_0^z
\frac{dz'}{H(z)}$.

In the next section, we will present numerical results to see the effect of the interaction between DE and DM. In each situation we will study
the cases with homogeneous and inhomogeneous DE distributions. When the DE is distributed inhomogeneously, it will participate the collapse and
the structure formation. We are not going to seek precise confrontations with observational data in this work, but to understand the influence of
coupling between dark sectors on cluster number counts.

\section{cluster number counts\label{numerical}}
In this section we present numerical results on the cluster number counts due to different interactions between DE and DM. We limit our
attention to three commonly used forms of interaction between dark sectors, with the interaction proportional to the energy densities of DE, DM
and total dark sectors respectively. In our numerical calculation, we consider that all models have the same $\sigma_8=0.8$ . In the end of this section we also
discuss the results by normalizing the halo abundance to the present value.

\subsection{Interaction proportional to the energy density of DE ($\xi_1=0,\xi_2 \neq 0$)}

\begin{figure}
\begin{center}
  \begin{tabular}{cc}
\includegraphics[width=3in,height=3in]{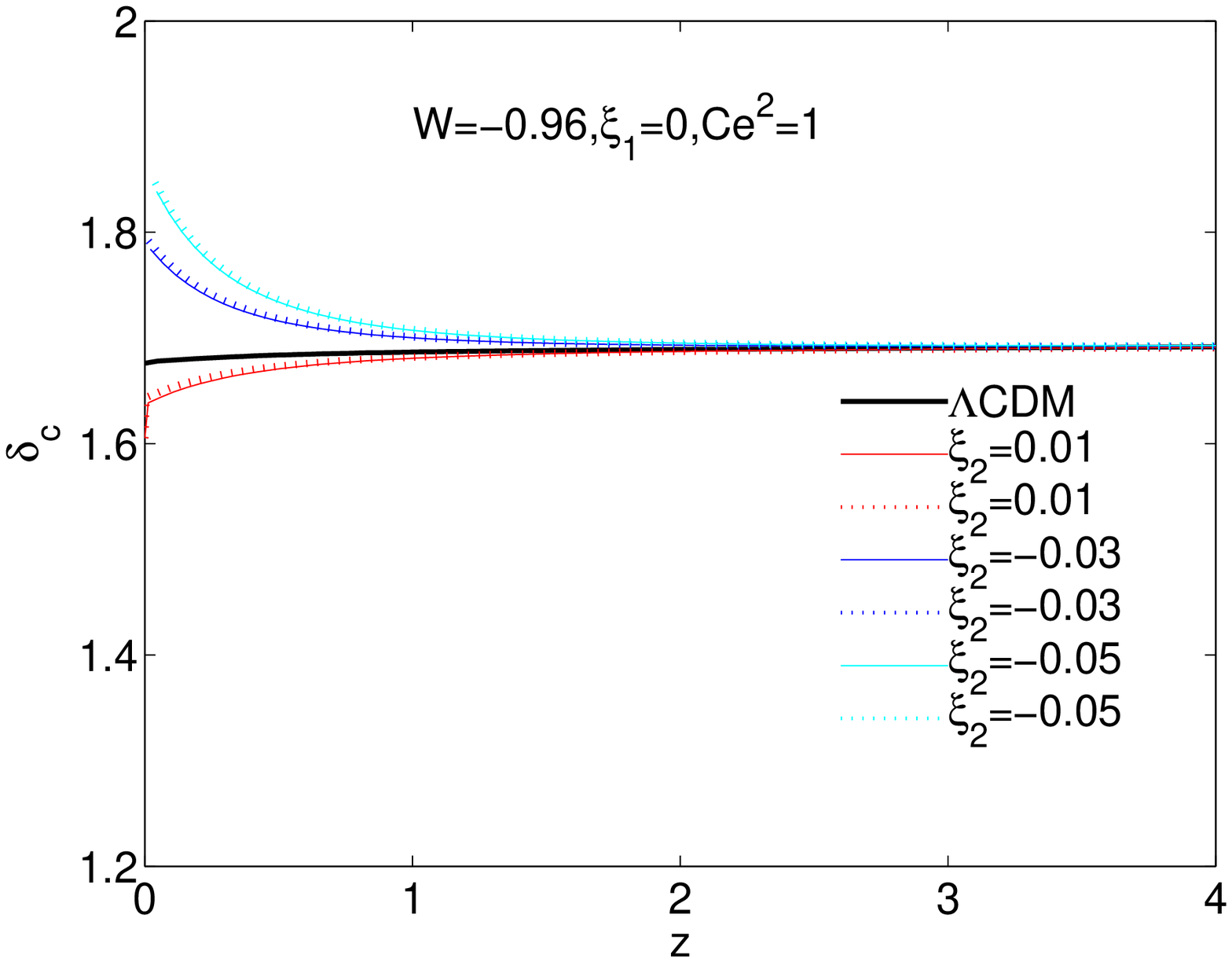}&
\includegraphics[width=3in,height=3in]{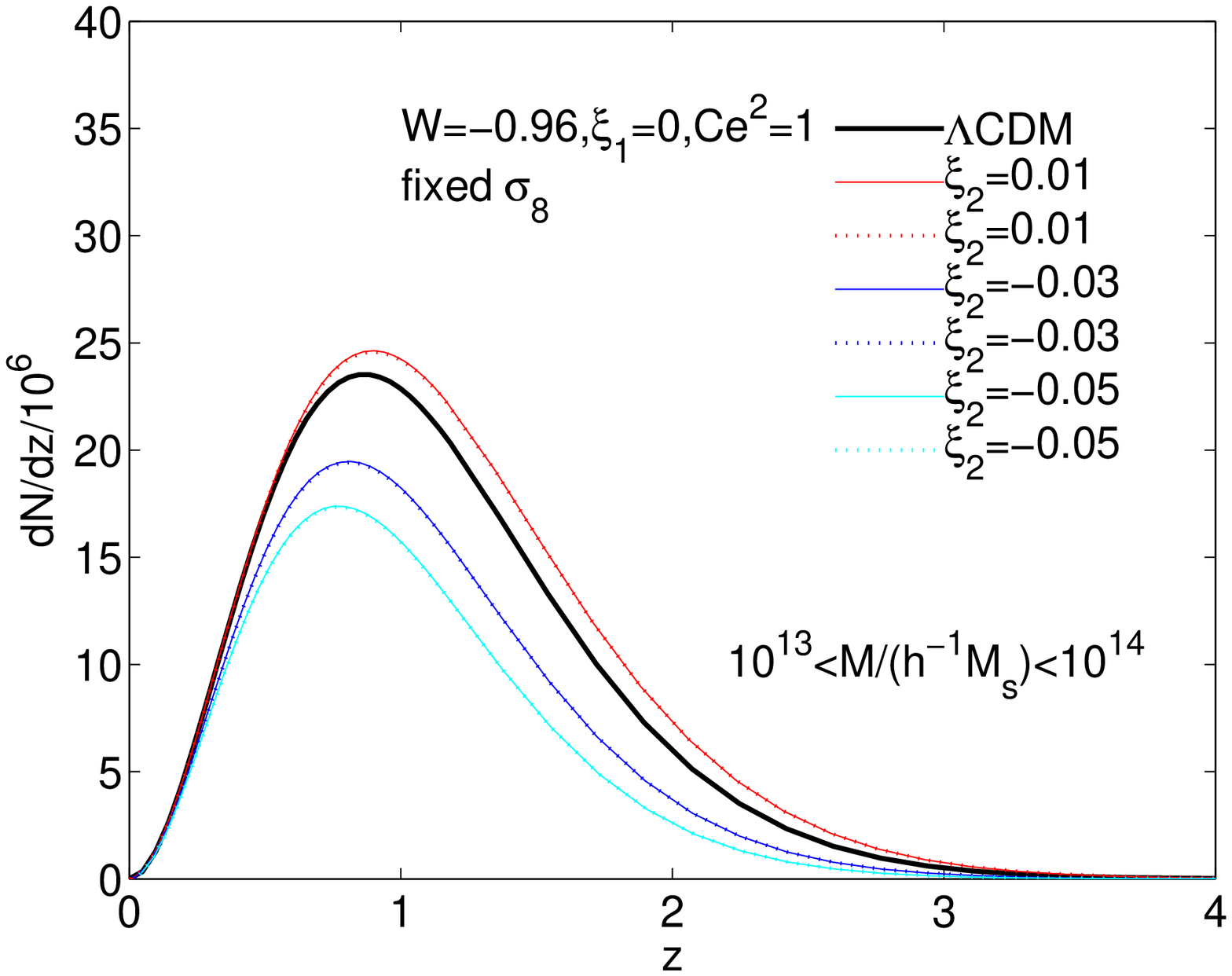} \\
    (a)&(b) \\
\includegraphics[width=3in,height=3in]{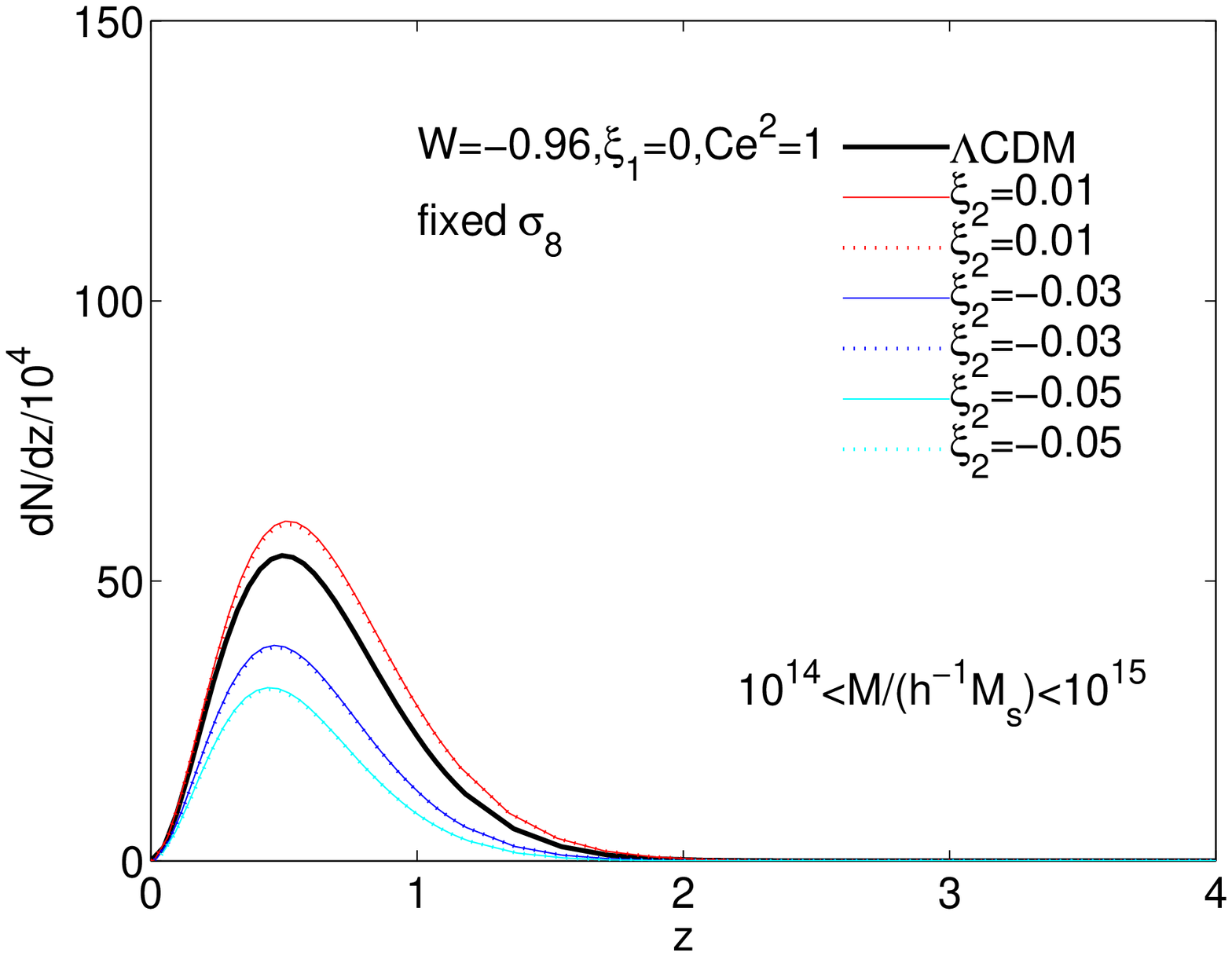}&\includegraphics[width=3in,height=3in]{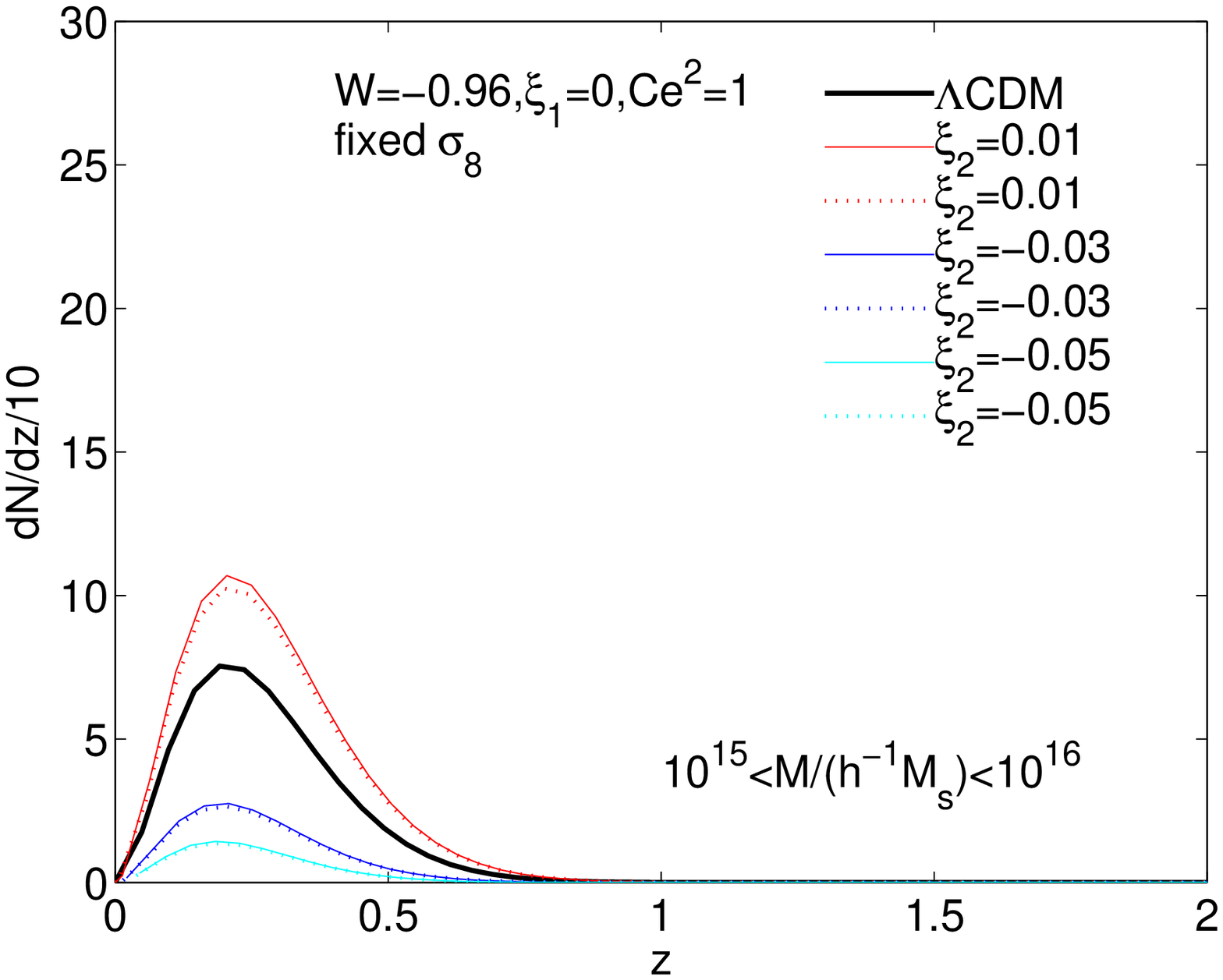}\\
    (c)&(d)\\
  \end{tabular}
\end{center}
\caption{The numerical results for the coupling proportional to the energy density of DE when the EOS of DE $w>-1$ . The solid lines show the
results calculated with homogeneous DE distribution and the dotted lines are for the inhomogeneous distribution of DE.}\label{dequint}
\end{figure}

When the interaction is proportional to the DE density, it was found that the curvature perturbation is always stable for both
quintessence and phantom DE EOS \cite{hePLB09}. When we consider DE EoS $w>-1$, we have the numerical results as shown in Fig~\ref{dequint}.

In the case when DE distribution is homogeneous, the results are shown in solid lines.  With positive coupling ($\xi_2>0$ DE decays to DM), we
see in Fig~\ref{dequint}a that the critical mass density decreases compared with the ${\rm \Lambda}$CDM model. This means that with the positive
coupling, cluster can be formed easier than that of ${\rm \Lambda}$CDM model. However when the coupling is negative, with DM decays to DE,
$\delta_c$ is higher than that of ${\rm \Lambda}$CDM, which means that it is harder for the cluster to be formed. The impact of the coupling between dark sectors is significant in causing the difference in the
cluster number counts. For positive coupling ($\xi_2>0$ DE decays to DM), the cluster number counts are bigger than that of ${\rm \Lambda}$CDM
model, which is clearly shown in Fig~\ref{dequint}b-Fig~\ref{dequint}d. However when the coupling is negative($\xi_2<0$ with DM decays to DE),
the situation is opposite and consequently cluster number counts are smaller than that of ${\rm \Lambda}$CDM model.

When the DE is distributed inhomogeneously, we need to consider the DE perturbation and its participation in the
structure formation. Results are shown in dotted lines in Fig~\ref{dequint}. We see that $\delta_c$ is slightly larger than the results without DE perturbation. The difference brought by the DE inhomogeneity is really
small if compared with that of the interaction.

\begin{figure}
\begin{center}
  \begin{tabular}{cc}
\includegraphics[width=3in,height=3in]{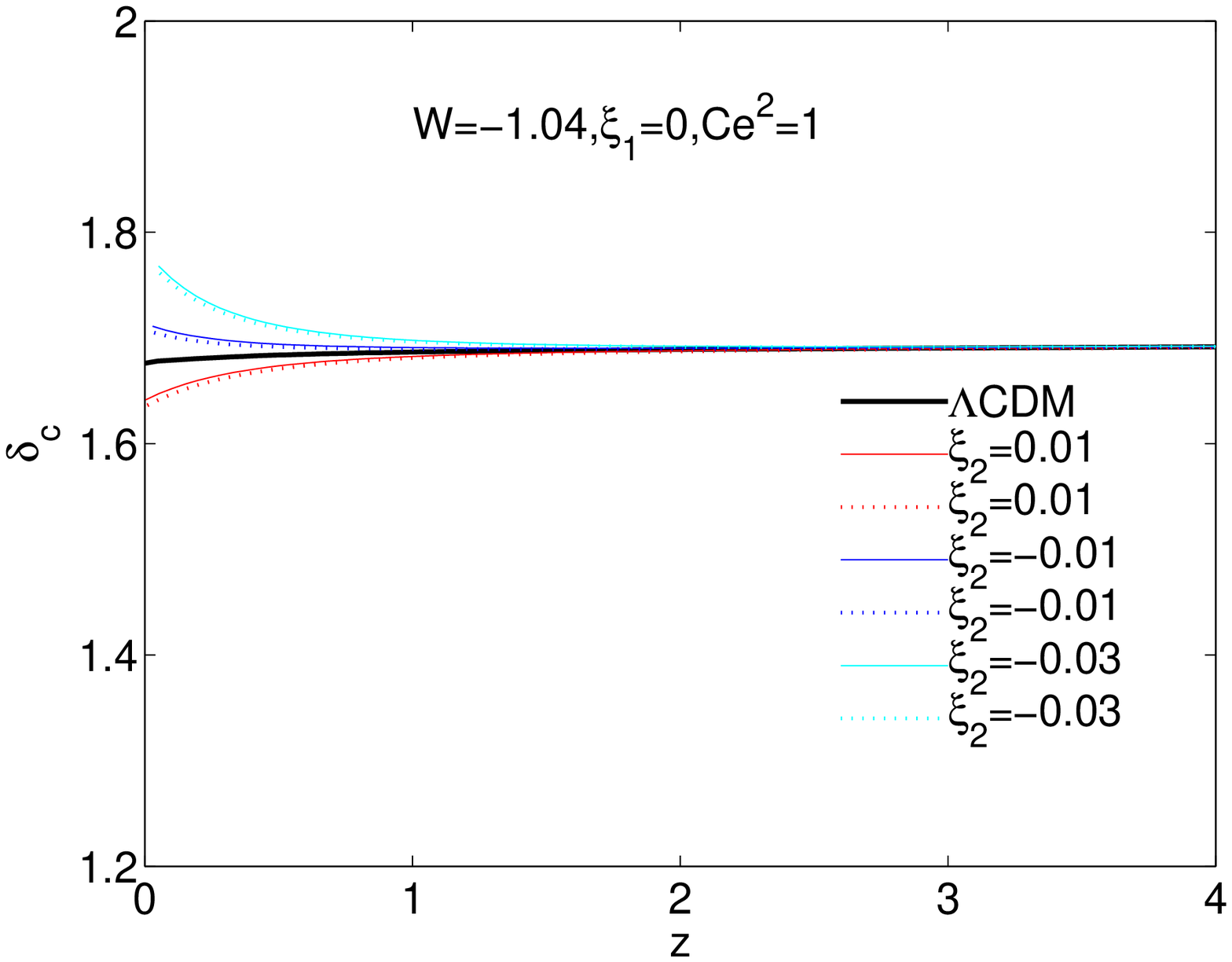}&
\includegraphics[width=3in,height=3in]{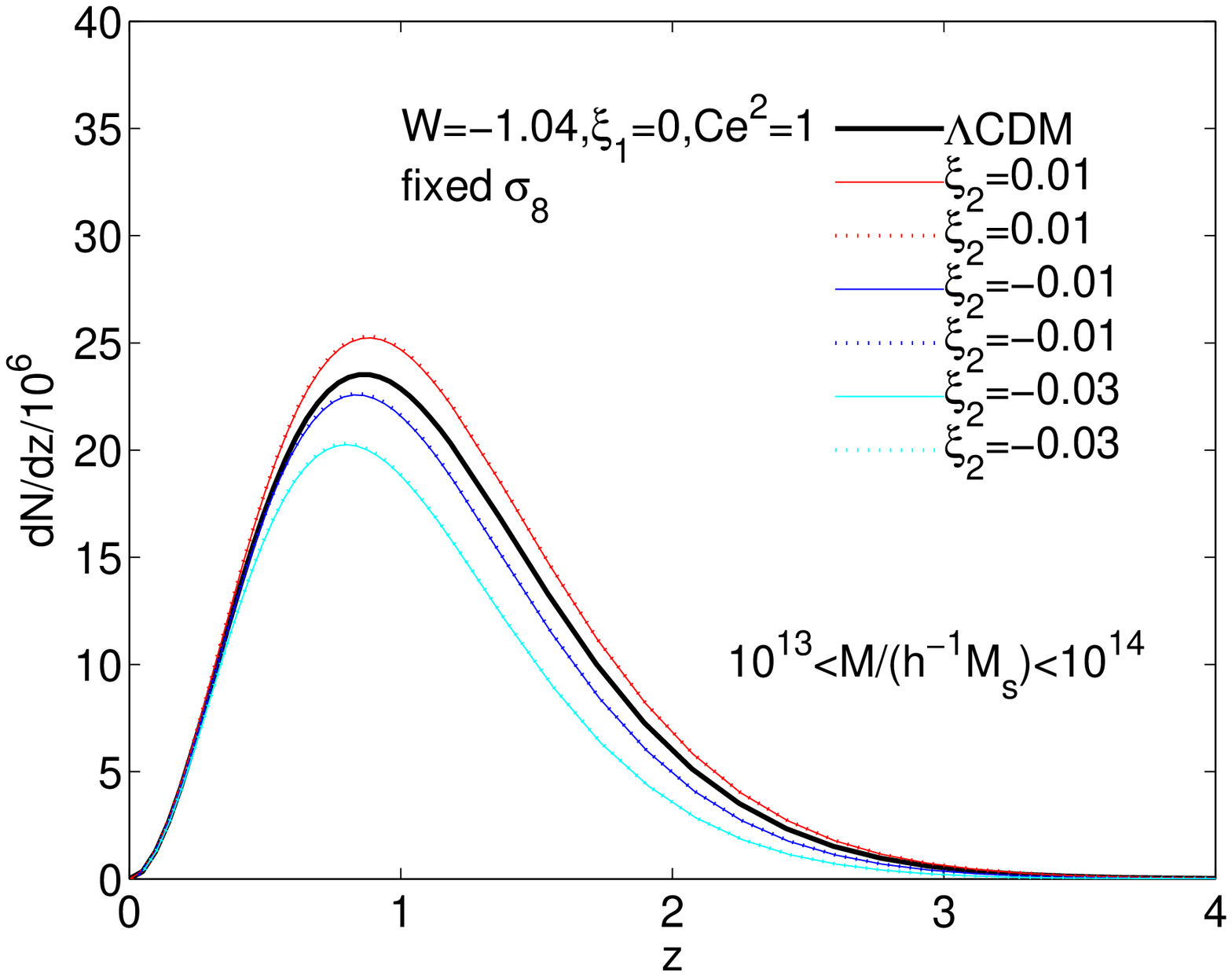}\nonumber \\
    (a)&(b)\nonumber \\
\includegraphics[width=3in,height=3in]{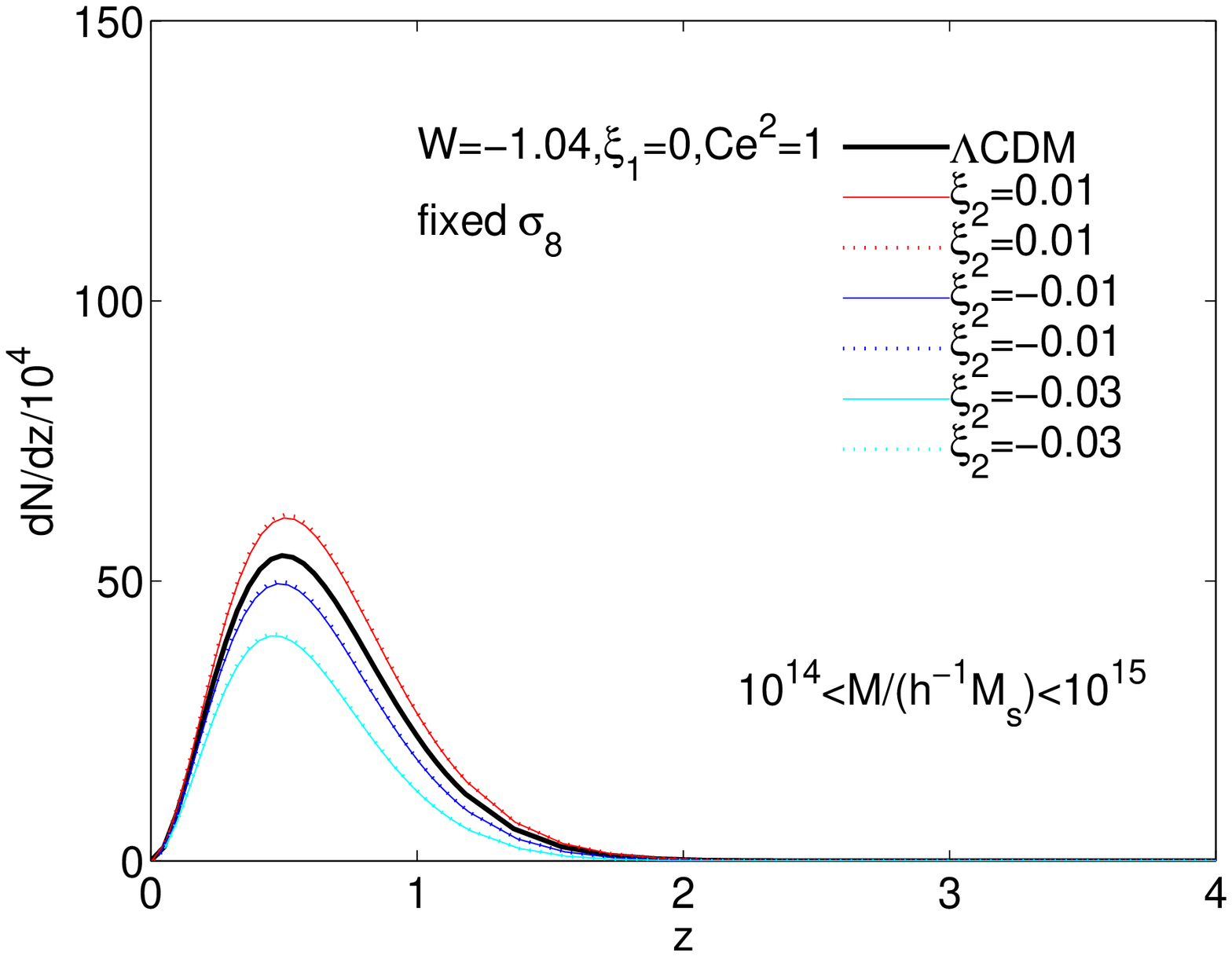}&\includegraphics[width=3in,height=3in]{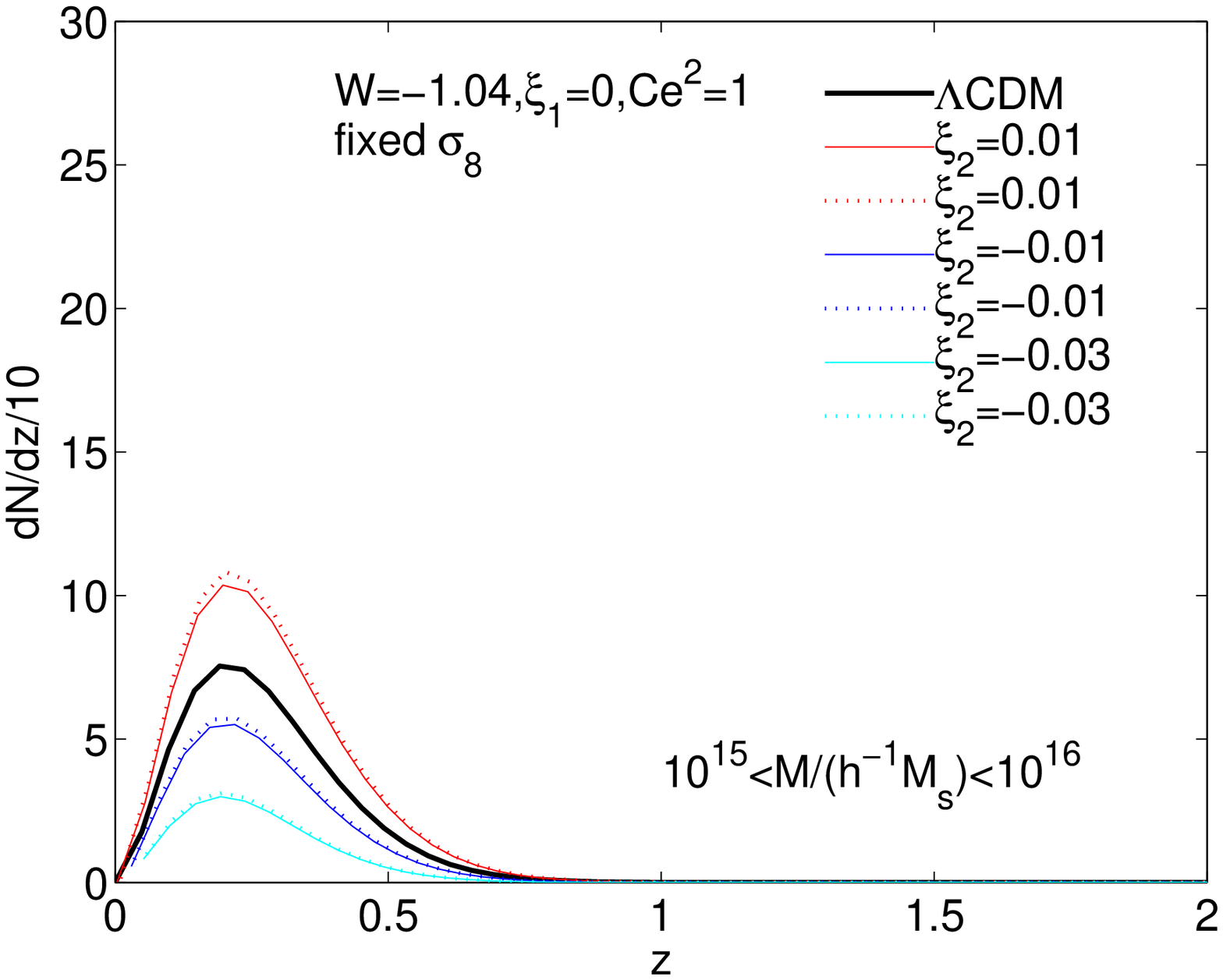} \\
    (c)&(d)\nonumber \\
   \end{tabular}
\end{center}
\caption{The numerical results for the coupling proportional to the energy density of DE when the EOS of DE $w<-1$ . The solid lines show the
results calculated when DE is distributed homogeneously and the dotted lines are for inhomogeneous DE model.}\label{dephantom}
\end{figure}

For the DE EoS $w<-1$, the numerical results are shown in Fig~\ref{dephantom}. When the DE distribution is homogeneous, the influence brought by
the interaction between dark sectors on the evolution of critical overdensity $\delta_c$ and the galaxy number counts agrees well with the
situation when $w>-1$. However when we consider the DE is distributed inhomogeneously, the model with clustering DE with
$w<-1$ has a bit more number counts than the homogeneous DE distribution appeared in the large mass bin,  although the difference is very small
especially for the case of small mass bin.

Comparing with the interaction, we see that the influence brought by the difference between the homogeneous and inhomogeneous DE distributions on
the critical density $\delta_c$ and the number of galaxy clusters is very small. This in fact tells us that the fluctuation of DE field
$\sigma_d$ is small. In eq.(\ref{DEv}), this small $\sigma_d$ leads to $U_{md}\sim U_{dm}\sim U_{dd}\sim T_{d}\rightarrow 0$, which means that
the DE plays very little role in the virialization of the structure. The structure formed is almost the same as that of the cluster with
homogeneous DE distribution. The little effect of the inhomogeneous DE distribution in the structure formation is consistent with the finding
that the DE perturbation is small compared with the DM perturbation in the subhorizon approximation \cite{31}.

\subsection{The interaction proportional to the energy density of DM ($\xi_1>0,\xi_2=0$)\label{DM}}

\begin{figure}
\begin{center}
  \begin{tabular}{cc}
\includegraphics[width=3in,height=3in]{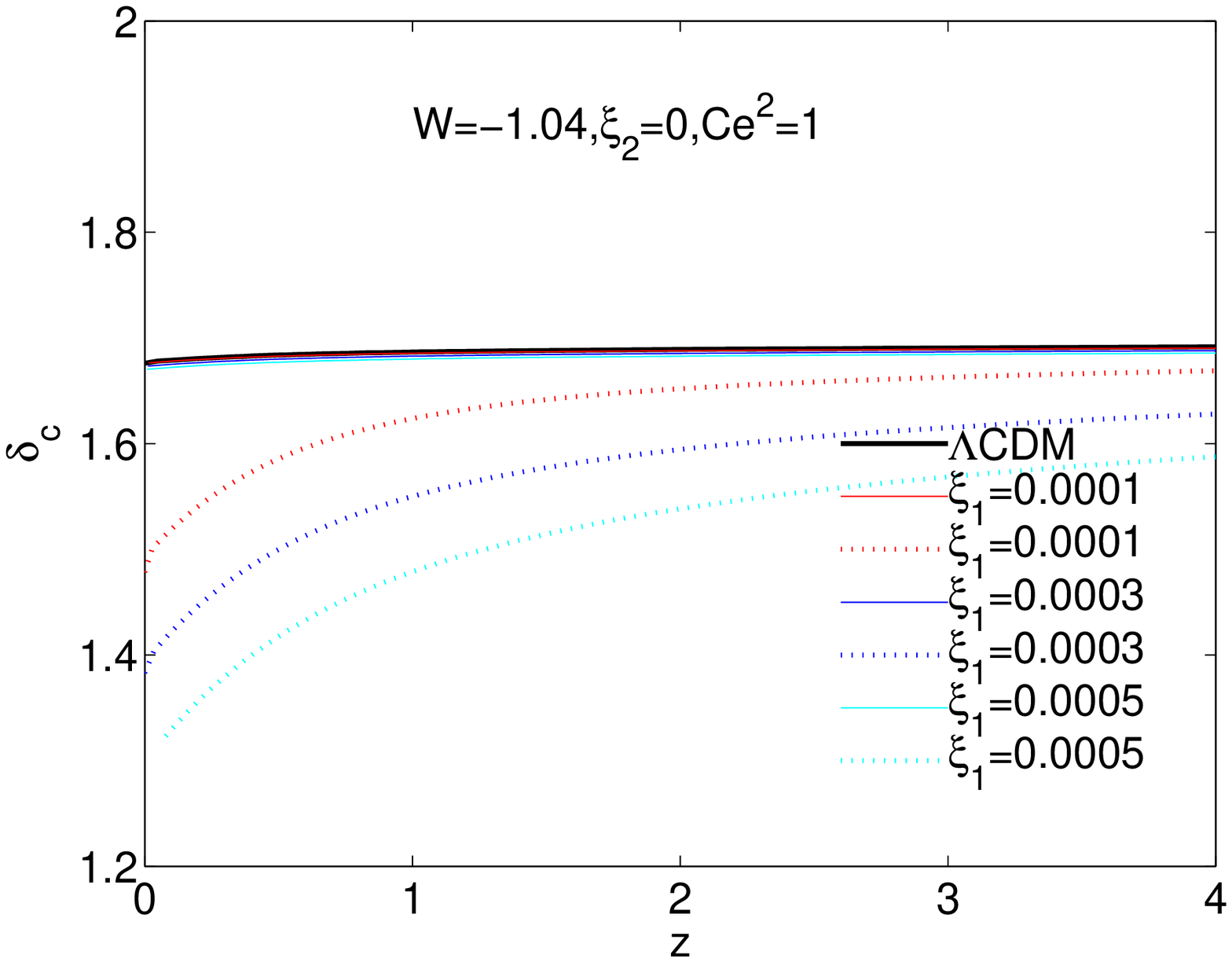}&
\includegraphics[width=3in,height=3in]{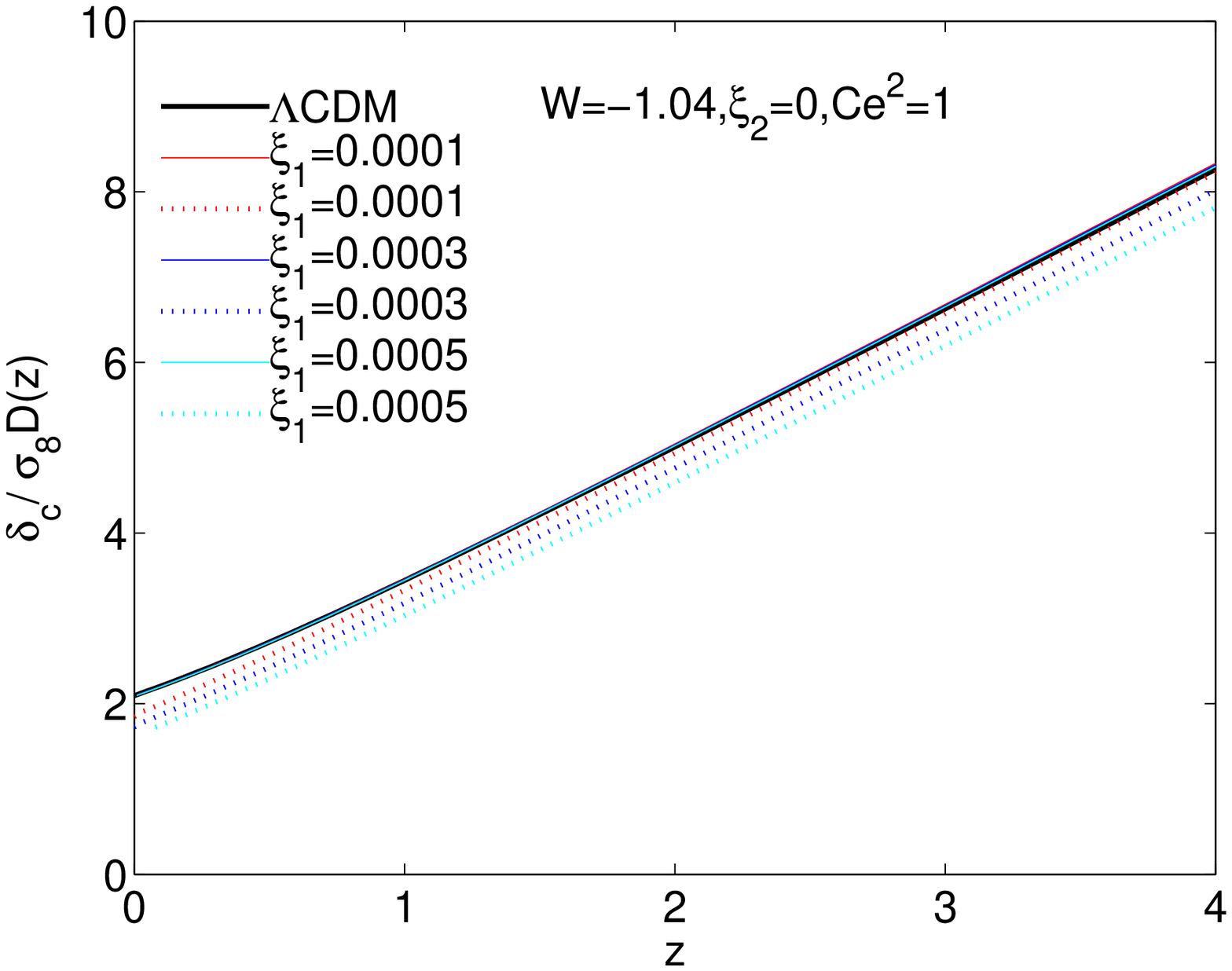}\nonumber \\
    (a)&(b)\nonumber \\
\includegraphics[width=3in,height=3in]{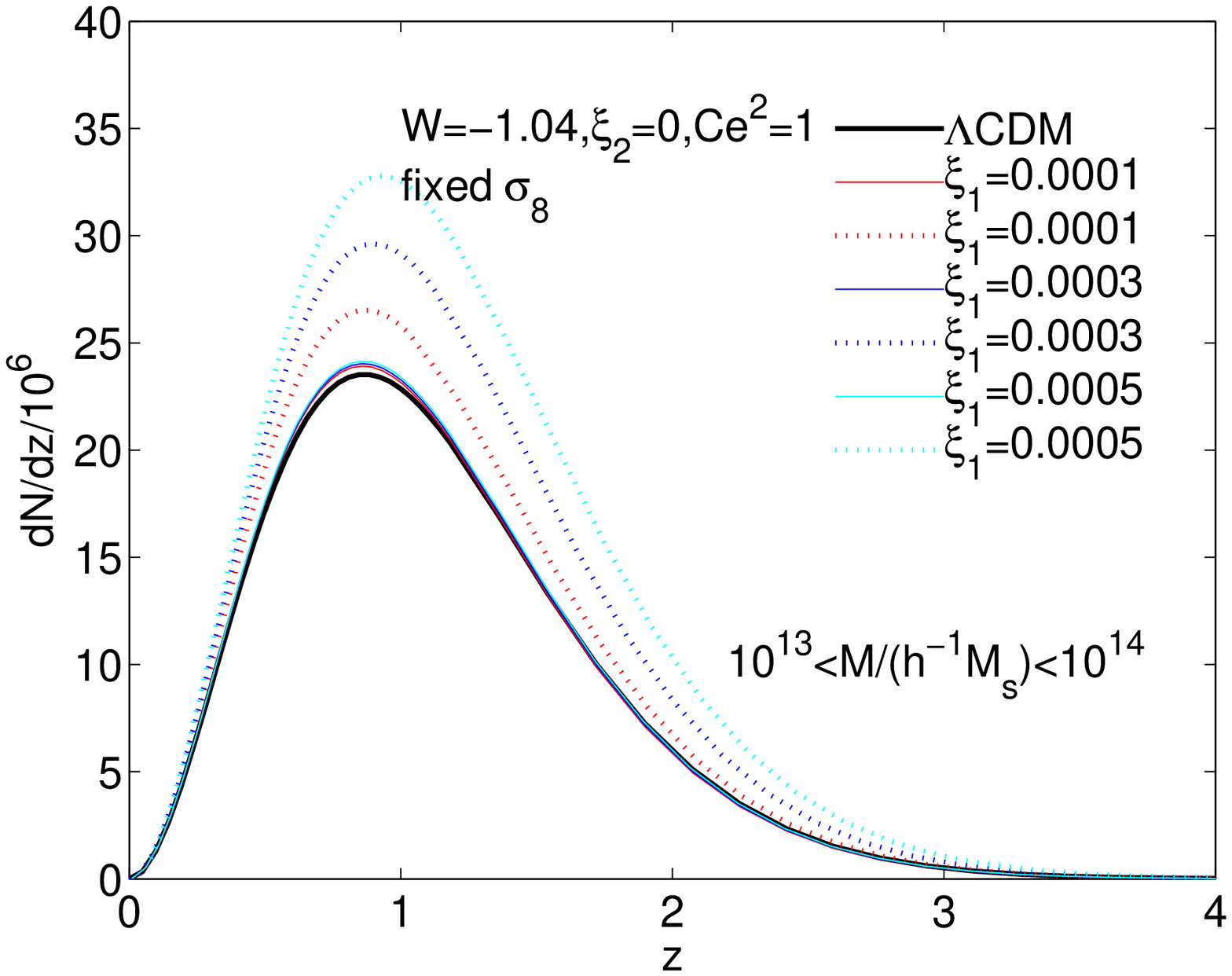}&\includegraphics[width=3in,height=3in]{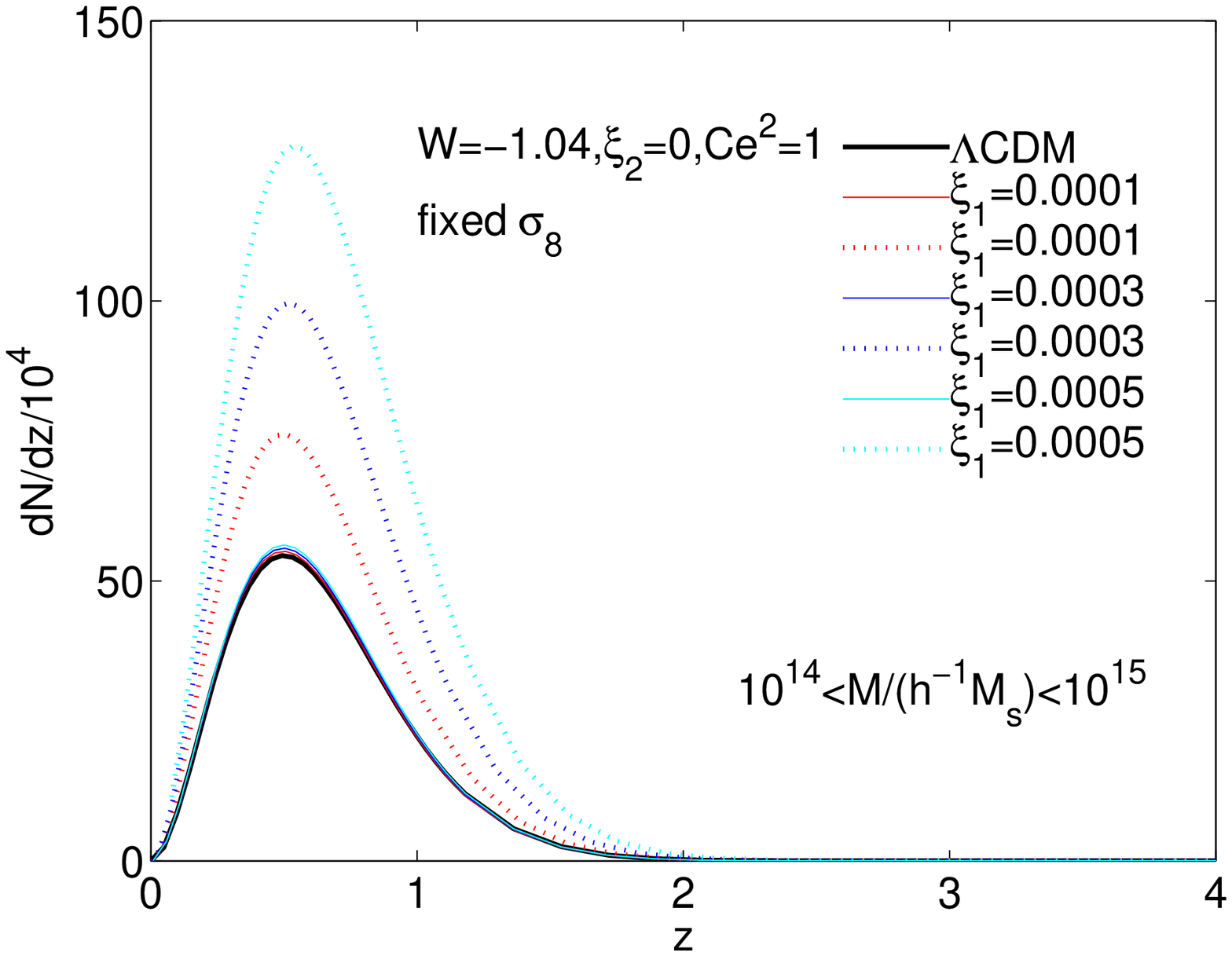}\\
    (c)&(d)\nonumber \\
  \end{tabular}
\end{center}
\caption{The numerical results for the coupling proportional to the energy density of DM. The solid lines shows the results
calculated with homogeneous DE distribution and the dotted lines are for inhomogeneous DE.}\label{dm}
\end{figure}

In this case, the stable curvature perturbation can only hold when DE EOS is smaller than -1\cite{hePLB09}. The coupling between DE and DM should
be positive to avoid the negative energy density of DE at the early time of the universe\cite{heJCAP08} The numerical results by choosing this
coupling are shown in Fig~\ref{dm}.

The solid lines are for the homogeneous distribution of DE. For small coupling, the lines are close to that of ${\rm \Lambda}$CDM result. We observe that more positive coupling
leads to more number counts of clusters Fig~\ref{dm}c. This can be understood from $\delta_c$ and also the ratio $\delta_c(z)/\sigma_8D(z)$,
which are smaller at high redshift for more positive $\xi_1$.  This result is also consistent with the case when the interaction is proportional
to the DE energy density.

The dotted lines are for the situations when DE is distributed inhomogeneously and we need to consider the DE perturbation and its participation
in the structure formation. From the behavior of $\delta_c$ we see that $\delta_c$ is suppressed at low redshift. This suppression is more
obvious than the case of homogeneous DE. In the ratio $\delta_c(z)/\sigma_8D(z)$ shown in Fig~\ref{dm}b, the ratio due to the inhomogeneous DE is
smaller than that of the homogeneous DE. This leads to a larger cluster abundance due to the inhomogeneous DE distribution than that of
homogeneous DE distribution as shown in Fig~\ref{dm}c.

\subsection{Interaction proportional to the total energy density of dark sectors  ($\xi=\xi_1=\xi_2>0$)}

\begin{figure}
\begin{center}
  \begin{tabular}{cc}
\includegraphics[width=3in,height=3in]{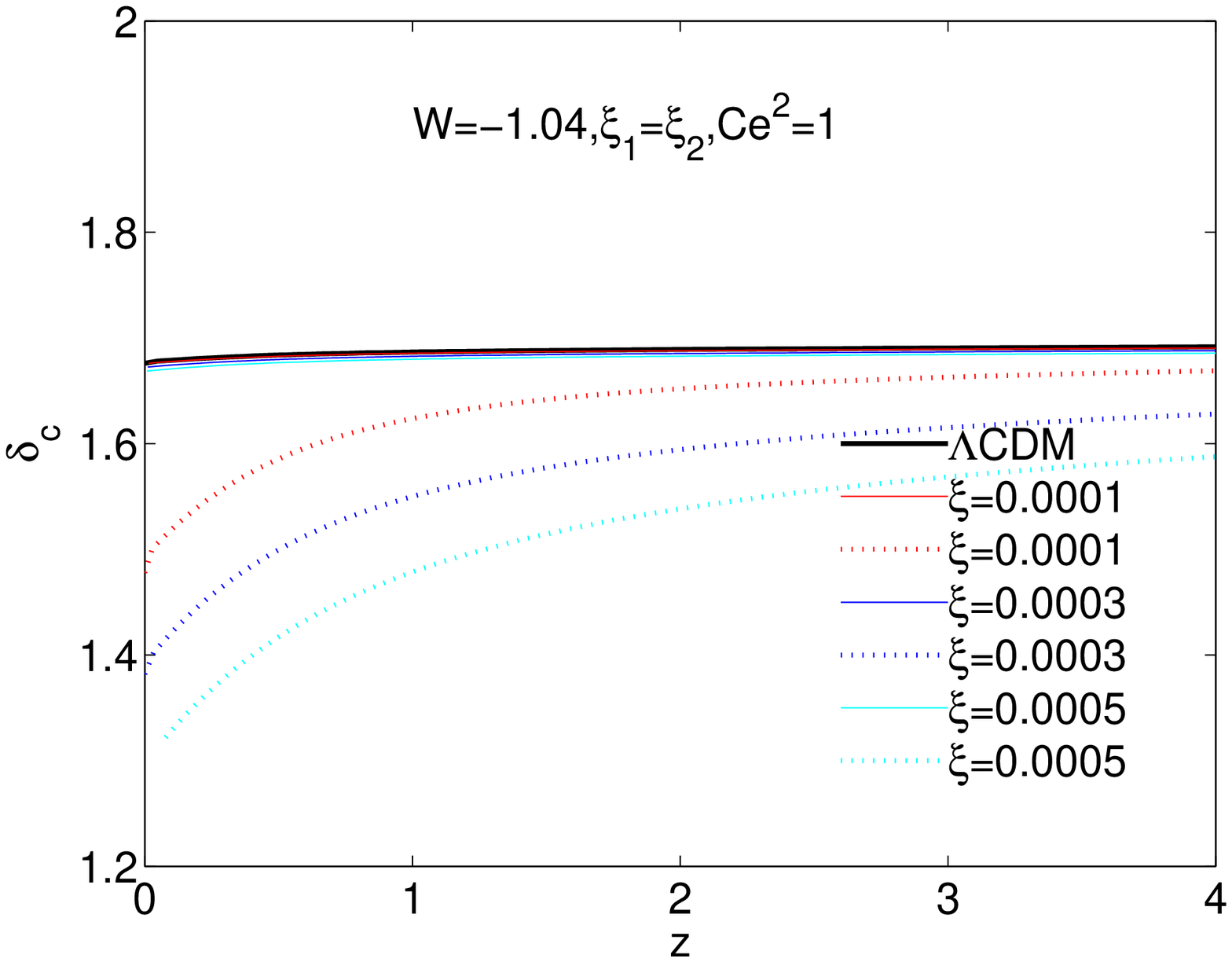}&
\includegraphics[width=3in,height=3in]{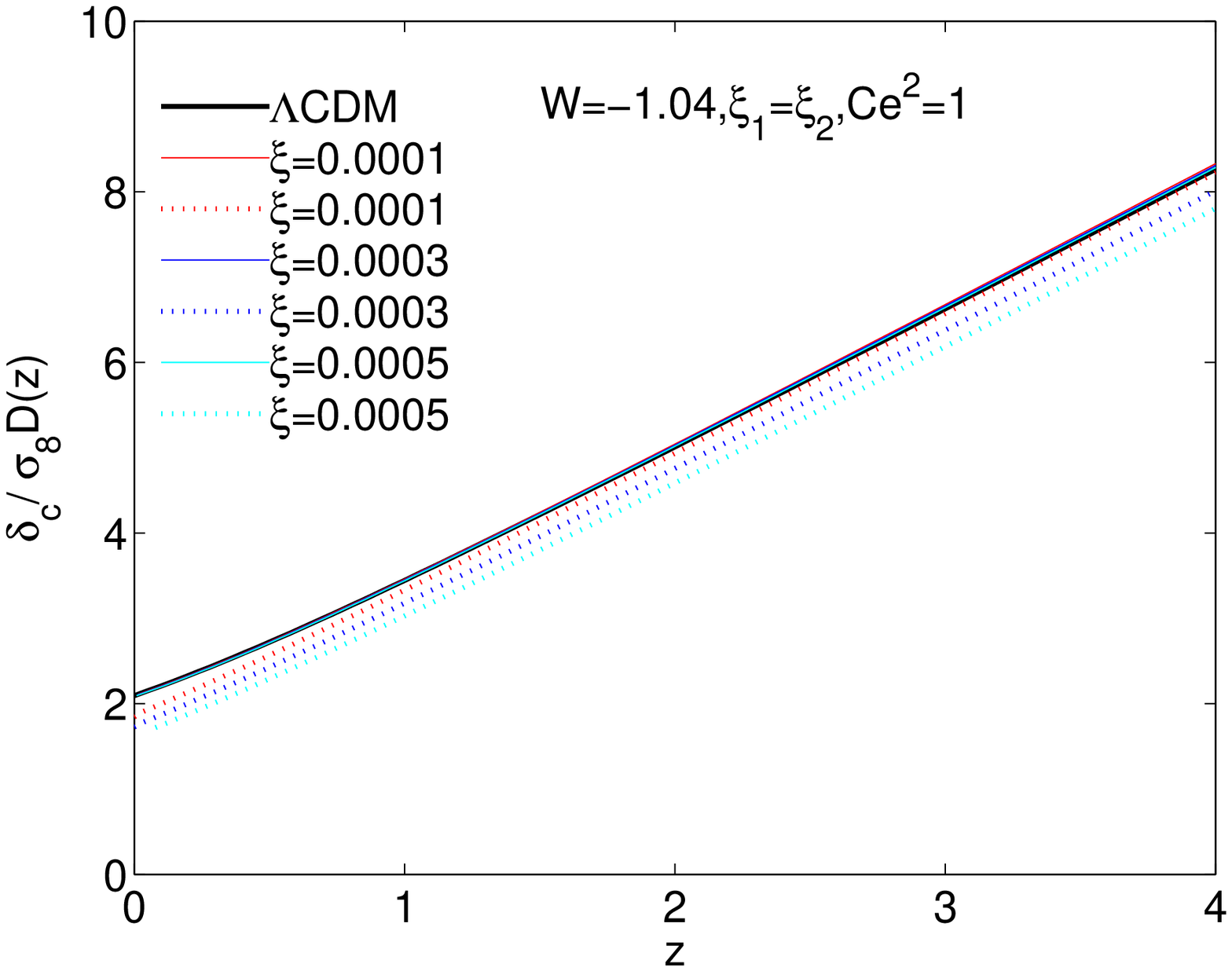}\nonumber \\
    (a)&(b)\nonumber \\
\includegraphics[width=3in,height=3in]{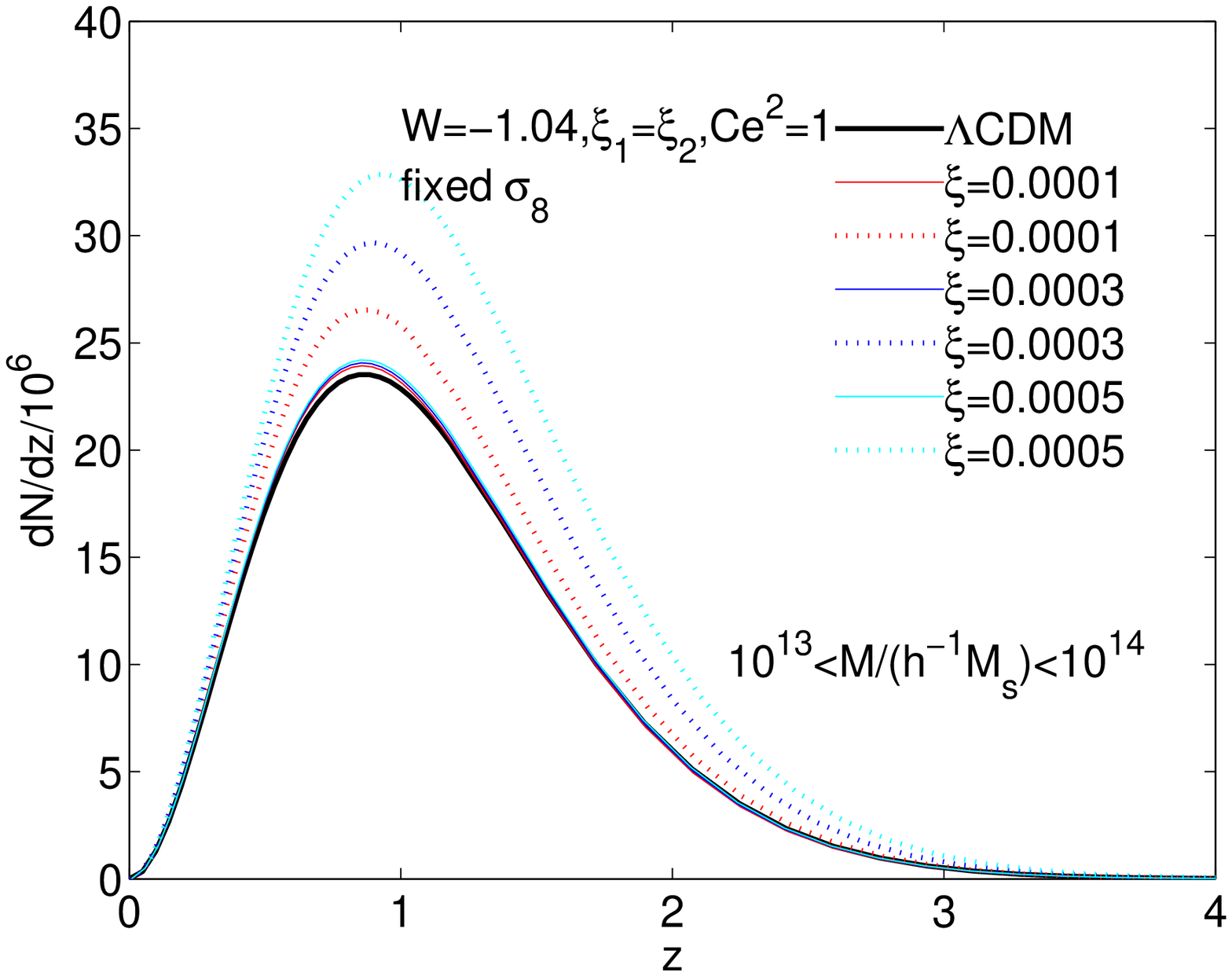}&\includegraphics[width=3in,height=3in]{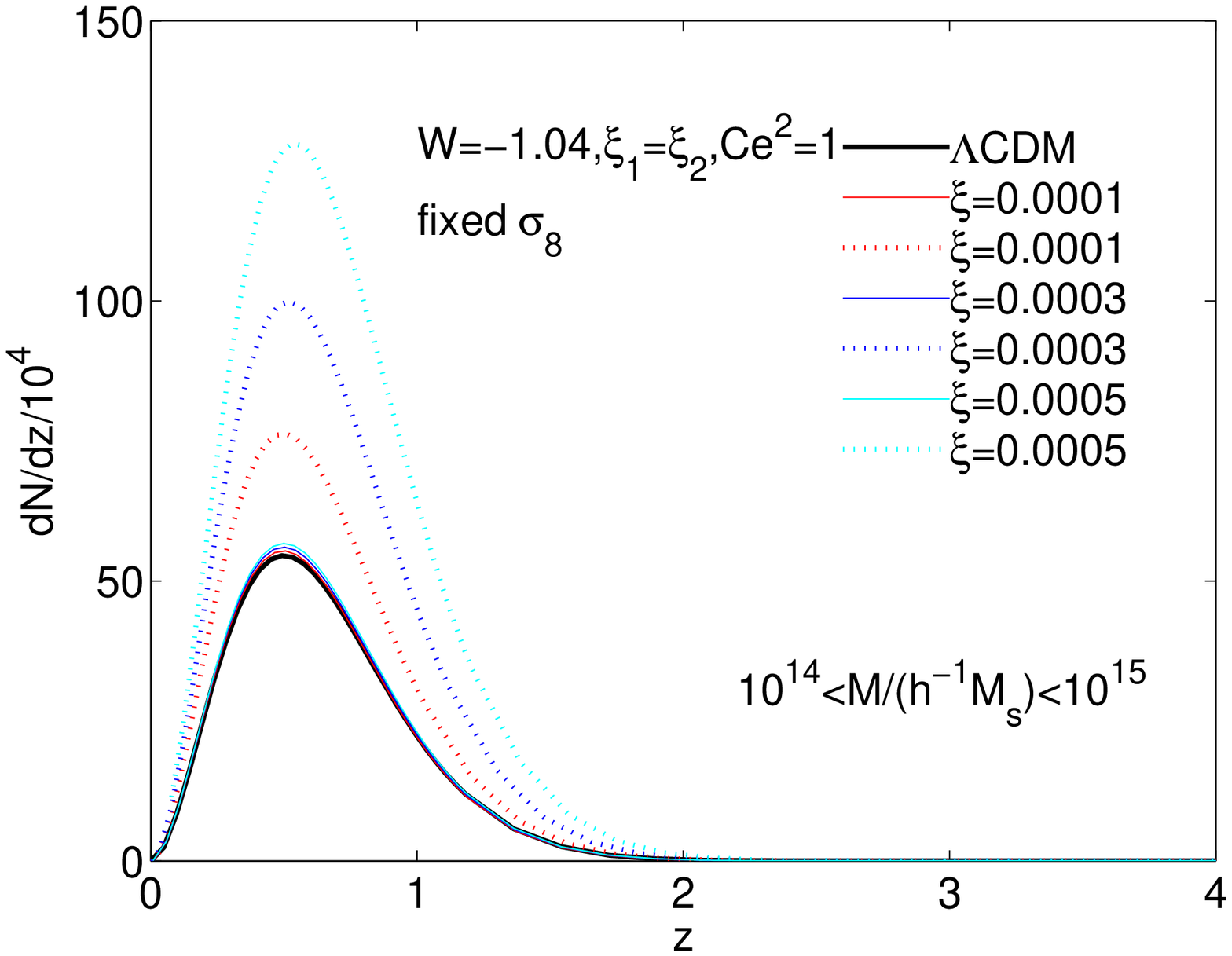}\\
 \end{tabular}
\end{center}
\caption{The numerical results for the coupling proportional to the energy density of total dark sectors. The solid lines shows the results
calculated with homogeneous DE distribution and the dotted lines are for inhomogeneous DE.}\label{t}
\end{figure}

Here, again, the curvature perturbation can only be stable when DE EOS is smaller than -1 \cite{hePLB09} and the coupling has to be positive
\cite{heJCAP08}. The numerical results are shown in Fig~\ref{t}. The results are similar to the case when the coupling is proportional to the
energy density of DM. With more positive coupling, the critical overdensity is smaller which leads to the bigger galaxy number counts than that
of the ${\rm \Lambda}$CDM model. For the same coupling, if the DE is clustering, the critical overdensity $\delta_c$ and
$\delta_c(z)/\sigma_8D(z)$ will be suppressed at small redshift. Therefore it is observed in Fig~\ref{t}c,Fig~\ref{t}d that at small the
inhomogeneous DE distribution leads to a larger cluster number abundance as compared with the homogeneous DE distribution.

\subsection{Normalization}

So far in the above discussions we have fixed $\sigma_8=0.8$ as given by present observations \cite{sperel}. Considering that at redshift zero
all models have the same number density of halos, the normalization is required to be done by adjusting $\sigma_8$ in each model such that
$\delta_c/\sigma_8$ is equal to the fiducial $\sigma_8=0.8$ ${\rm \Lambda}$CDM model
\begin{equation}
\sigma_{8,model}^0=\frac{\delta_{c,model}(z=0)}{\delta_{c,\Lambda}(z=0)}
\sigma_{8,\Lambda}^0\quad . \label{nomal}
\end{equation}
We dedicate the rest of this section to discuss the implications on our results by normalizing models to the same halo abundance at $z=0$. It is
interestingly observed that the qualitative structure of curves  changes dramatically depending on the choices of the normalizations, which was
also found in studying the number counts in homogeneous and inhomogeneous DE models without coupling to DM \cite{numes2005}.

\begin{figure}
\begin{center}
\begin{tabular}{cc}
\includegraphics[width=3in,height=3in]{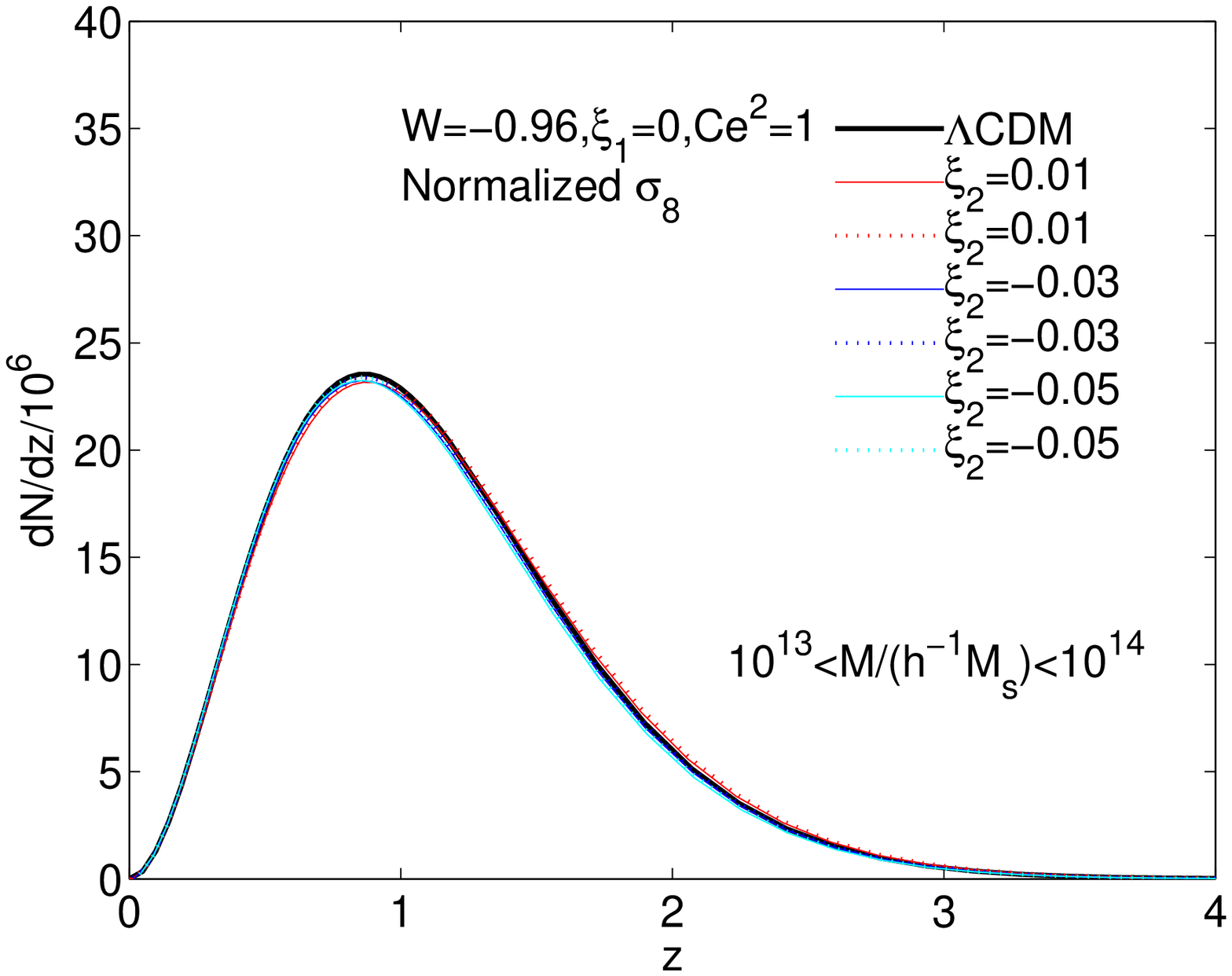} &
\includegraphics[width=3in,height=3in]{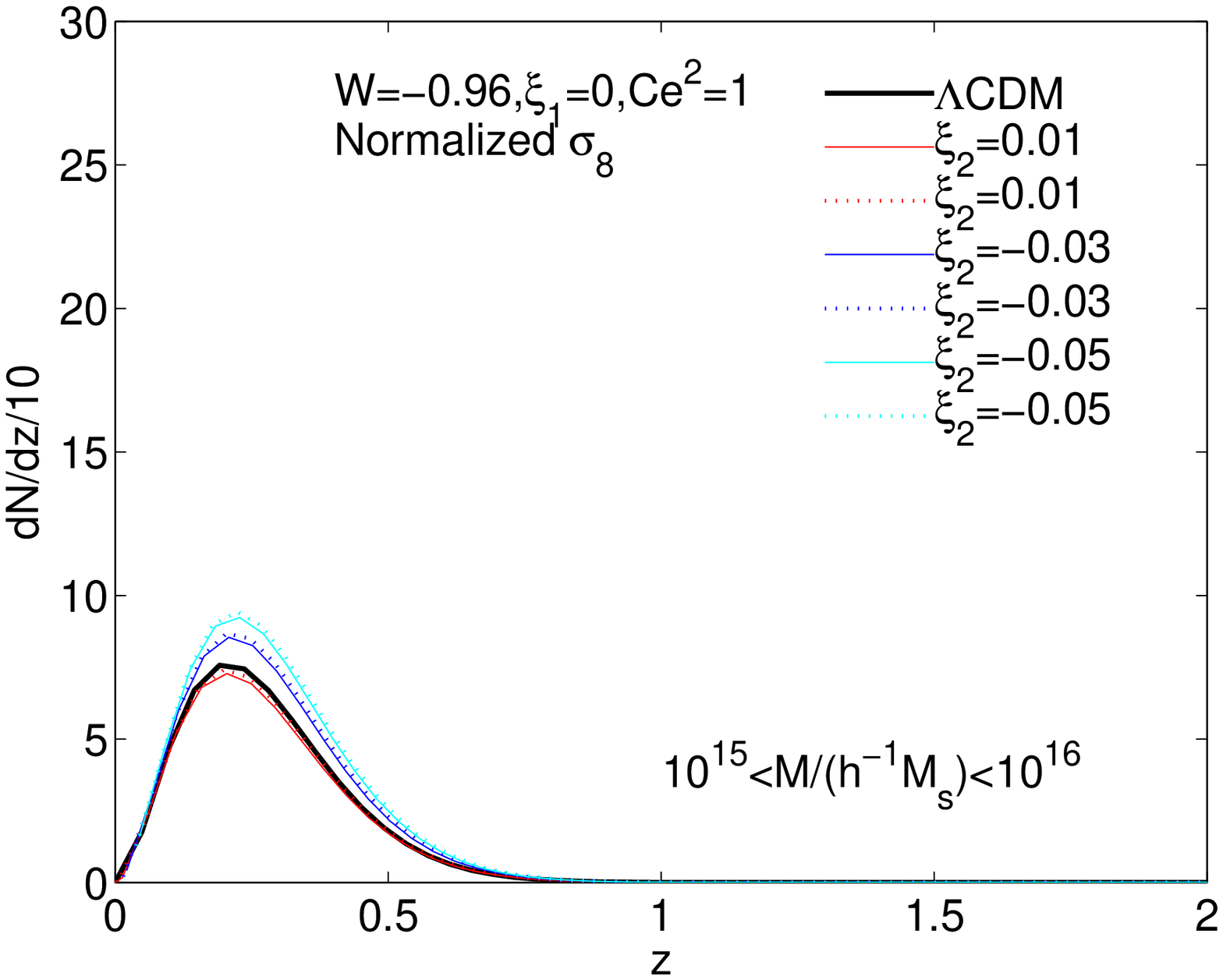} \nonumber \\
(a) & (b) \nonumber \\
\end{tabular}
\end{center}
\caption{ The results for the coupling proportional to the energy density of DE by fixing the local abundance.  }
\label{normalde}
\end{figure}

In our models, we find in Fig~\ref{normalde}a that when the coupling is proportional to the energy density of DE($\xi_1=0,\xi_2\neq 0$), in the
normalization of local halos abundance, the departure from the cosmological constant model due to the interaction between dark sectors becomes
smaller especially in the low mass bin when compared with the results by fixing $\sigma_8$. The major impact due to the coupling emerges in the
massive galaxy cluster as shown in Fig~\ref{normalde}b. The signature of the
 DE perturbation is still not obvious which is similar to the result by fixing $\sigma_8$.

\begin{figure}
\begin{center}
\begin{tabular}{cc}
\includegraphics[width=3in,height=3in]{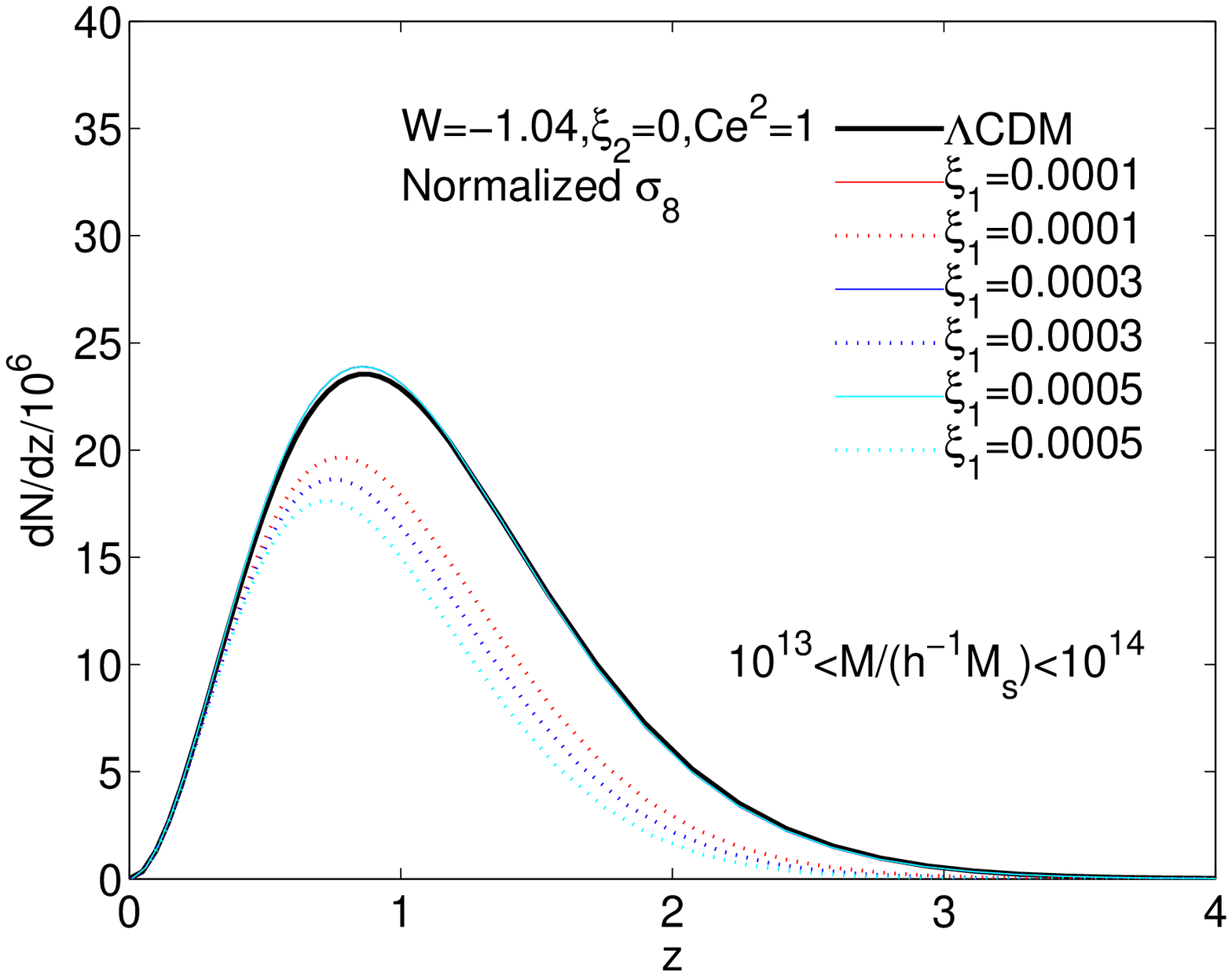} &
\includegraphics[width=3in,height=3in]{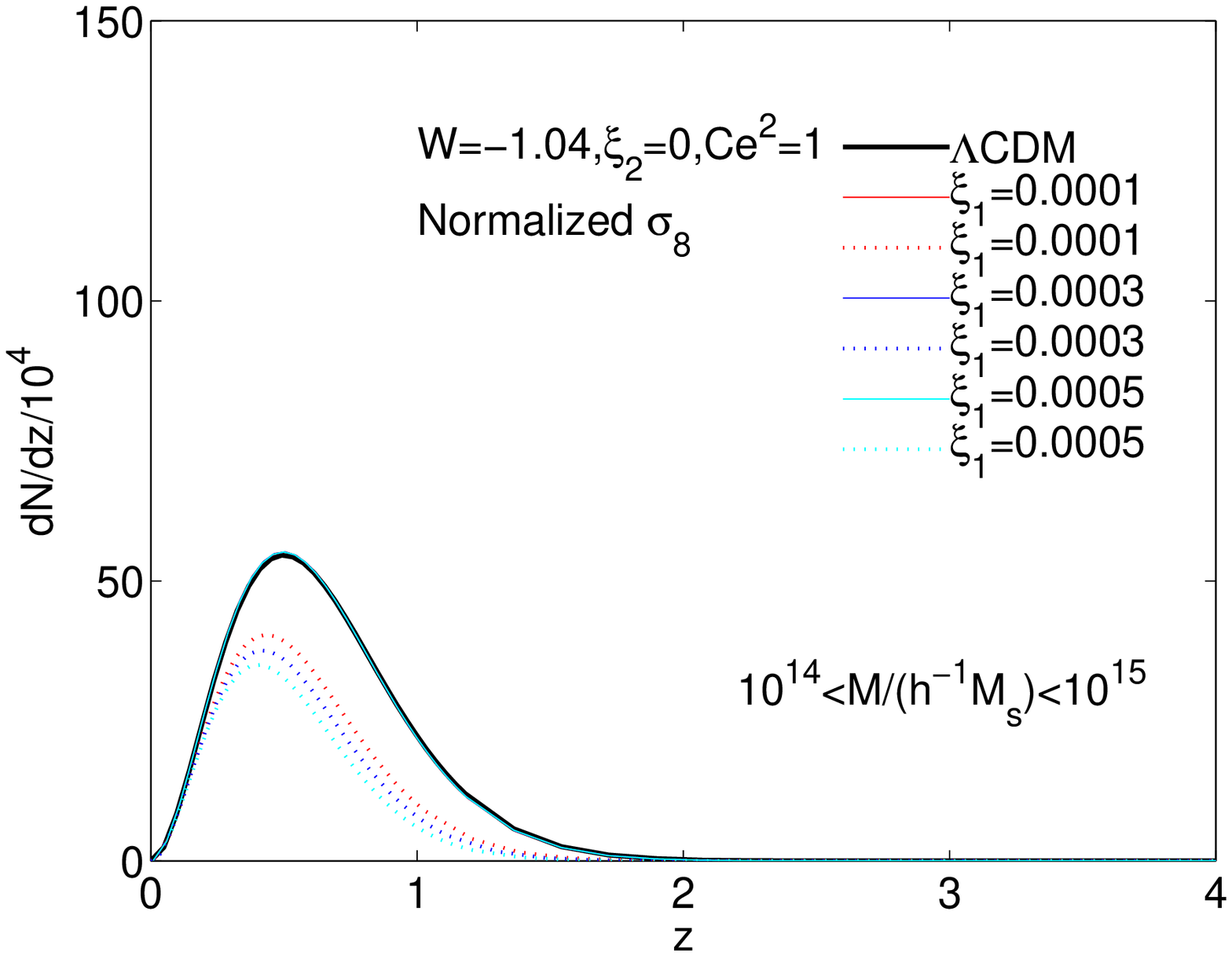} \nonumber \\
(a) & (b)  \nonumber \\
\end{tabular}
\end{center}
\caption{ The results for the coupling proportional to the energy density of DM by fixing the local abundance. } \label{normaldm}
\end{figure}

When the coupling is proportional to the energy density of DM, we see from Fig~\ref{normaldm} that the departure from the cosmological constant
model due to the interaction between DE and DM (solid lines) is suppressed if compared with the result by fixing $\sigma_8$ Fig~\ref{dm}. This shows that fixing the
halo abundance, the coupling does not show up in the galaxy number counts. However, in this case, the difference caused by the clustering DE (dotted lines) is
enhanced with large deviations from the cosmological constant model. When the coupling is proportional to the energy density of total dark
sectors, the qualitative results are the same.

\section{conclusions and discussions\label{conclusion}}

Assuming first that DE component is homogeneous, we have investigated how DM  approaches the dynamical equilibrium in the collapsing system. In
the presence of coupling between DE and DM, the flow of mass and energy between the components changes the time required by the system to reach
equilibrium and also the equilibrium configuration itself. Our discussion is based on the perturbation theory developed in \cite{hePLB09,31}. We
reproduce the Layzer-Irvine equation \cite{Layzer} when the interaction is neglected. In the homogeneous case, the DE component flows progressive
out of the overdensity which leads the energy conservation law invalid in the collapsing system \cite{greek}. It is expected in the models of DE
interacting with DM, if the DE is distributed inhomogeneously, it may cluster along with the DM and thus we can avoid the energy non-conservation
problem. To see whether this speculation is correct or not, we have further explored the dynamics of the DE that presents inhomogeneities at
cluster scales and found that in the presence of the coupling between DE and DM, the time and dynamics required by the DE in the system to reach
equilibrium is different from that of DM. This shows that in the collapsing system, DE does not fully cluster along with DM, even in the presence
of the interaction between them. The energy is not strictly conserved in the collapsing system. This result is based on the linear perturbation
formalism, where the DE perturbation can be neglected in the late universe [19].

We also developed a novel  treatment of the spherical collapse model which consists of multi-fluids with different collapsing velocities. When DE
does not trace DM in the spherical collapse,  we studied the dependence of the cluster number counts on the coupling between DE and DM and on the
DE inhomogeneities by using the Press-Schechter formalism. In most of our work we have assumed that the models are normalized to have practically
the same $\sigma_8=0.8$ as given by present day observations. We have shown that when the coupling between dark sectors is proportional to the
energy density of DE, there is a significant dependence of cluster number counts on the interaction. If the energy decays from DE to DM, the
number of cluster counts increases. The dependence of cluster number counts on the DE inhomogeneities is negligible compared with the interaction
influence. This in fact shows that the fluctuation of the DE field is small and the inhomogeneous DE plays little role in the virialization of
the structure. For the coupling between DE and DM proportional to the energy density of DM and total dark sectors, we have observed that more
energy decay from DE to DM leads more abundance of the cluster number counts compared with the ${\rm \Lambda}$CDM model. In addition, the
signature of the DE inhomogeneities is enhanced. This enhancement can be understood from the dynamics in which it shows that the DM gravitational
effect pulls the inhomogeneous DE to move and collapse along with DM.

We have further studied these models enforcing the normalization to reproduce the same abundance of galaxy clusters at redshift zero. The qualitative
structure changes in choosing different normalizations \cite{numes2005}. Comparing with the interaction effects in fixing $\sigma_8$,
we have verified that the departures from the ${\rm \Lambda}$CDM model are suppressed, the coupling between dark sectors while fixing the
local abundance usually does not show up in the galaxy number counts, except when the interaction is proportional to the energy density of DE in
the super mass bin. In fixing the local abundance, the influence caused by the inhomogeneous DE is enhanced with large deviations from the
cosmological constant model when the coupling is proportional to the energy density of DM and total dark sectors.

The Press-Schechter formalism we employed is good enough for the purpose of this paper to illustrate how the interaction between dark sectors
and the DE inhomogeneities influence cluster number counts, but it is still not easy to seek precise confrontations with observational data at
the present moment. Recently in the uncoupled model, the statistical analysis was performed on the sensitivity of cluster counts to DE
perturbations by employing a set of cosmological parameters using the Fisher matrix method. It was argued that the impact of DE fluctuations is
large enough, which could already be detected by existing instruments\cite{Abramo20072009}. It is of great interest to extend their analysis to
examine the interaction between DE and DM effect and the DE inhomogeneities influence. Such a study is under way.

\acknowledgements{ This work has been supported partially by NNSF of China and the National Basic Research Program of China under grant
2010CB833000 and brazilian foundations CNPq and FAPESP.}

\end{document}